\documentclass[apj,twocolumn,twocolappendix,numberedappendix,appendixfloats]{openjournal}

\usepackage{graphicx}
\usepackage{amsmath}
\usepackage{url}
\usepackage[breaklinks,colorlinks,citecolor=blue,linkcolor=blue,urlcolor=blue]{hyperref}
\providecommand{\doi}[1]{doi: \href{https://doi.org/#1}{\nolinkurl{#1}}}

\def\apjl{ApJ}%
\def\apjs{ApJS}%
\def\ao{Appl.~Opt.}%
\def\aap{A\&A}%
\newcommand{\jwst}{\textit{JWST}}
\newcommand{\targ}{GNz7q}
\newcommand{\ent}{GNz7q-ENT}

\providecommand{\farcs}{\mbox{$.\!\!^{\prime\prime}$}}

\begin{document}

\title{A tidal disruption event in a quasar at redshift 7.19
\vspace{-15mm}}
\shorttitle{A tidal disruption event in a quasar at $z=7.19$}
\shortauthors{Fujimoto et al.}

\author{
S.~Fujimoto$^{1,2}$\footnotemark[*],
 Q.~Fei$^{1,2}$,
 G.~B.~Brammer$^{3,4}$,
 K.~Inayoshi$^{5,6}$,
 V.~Kokorev$^{7,8}$,
 L.~C.~Ho$^{5,6}$,
 R.~Li$^{9}$,
 F.~Walter$^{3,10,11}$,
 V.~Bromm$^{7,8}$,
 L.~Colina$^{12}$,
P.~Dayal$^{1,13,14}$,
S.~L.~Finkelstein$^{7,8}$,
M.~Ginolfi$^{15}$,
Z.~Liu$^{16}$,
G.~C.~K.~Leung$^{16}$,
G.~E.~Magdis$^{3,4,17}$,
J.~Matthee$^{18}$,
R.~P.~Naidu$^{16}$,
P.~Oesch$^{19}$,
M.~Onoue$^{20,21}$,
P.~G.~P\'erez-Gonz\'alez$^{12}$,
D.~Watson$^{3,4}$,
J.~\'Alvarez-M\'arquez$^{12}$,
J.~Antwi-Danso$^{1,2}$,
Y.~Asada$^{1,2}$,
G.~Barro$^{12}$,
C.~M.~Casey$^{22}$,
J.~Chisholm$^{7,8}$,
S.~Gillman$^{3,17}$,
R.~Marques-Chaves$^{19}$,
J.~P.~U.~Fynbo$^{3,4}$,
T.~R.~Greve$^{3,4,17}$,
I.~Labbe$^{23}$,
R.~A.~Meyer$^{19}$,
F.~Rizzo$^{24}$,
A.~Robbins$^{25}$,
R.~Schneider$^{26,27}$,
C.~L.~Steinhardt$^{28}$,
S.~Toft$^{3,4}$,
A.~Trinca$^{26,29}$,
F.~Valentino$^{3,4,17}$,
R.~Valiante$^{26}$,
M.~Vestergaard$^{4,30}$,
and M.~Xiao$^{19}$
}

\affiliation{$^{1}$David A. Dunlap Department of Astronomy and Astrophysics, University of Toronto, 50 St. George Street, Toronto, Ontario, Canada}
\affiliation{$^{2}$Dunlap Institute for Astronomy and Astrophysics, University of Toronto, 50 St. George Street, Toronto, Ontario M5S 3H4, Canada}
\affiliation{$^{3}$Cosmic Dawn Center (DAWN), Copenhagen, Denmark}
\affiliation{$^{4}$Niels Bohr Institute, University of Copenhagen, Jagtvej 128, DK-2200 Copenhagen N, Denmark}
\affiliation{$^{5}$ Kavli Institute for Astronomy and Astrophysics, Peking University, Beijing 100871, China}
\affiliation{$^{6}$Department of Astronomy, School of Physics, Peking University, Beijing 100871, China}
\affiliation{$^{7}$Department of Astronomy, The University of Texas at Austin, Austin, TX 78712, USA}
\affiliation{$^{8}$Cosmic Frontier Center, The University of Texas at Austin, Austin, TX, USA}
\affiliation{$^{9}$Max-Planck-Institut f\"ur extraterrestrische Physik, Gie{\ss}enbachstra{\ss}e 1, 85748 Garching bei M\"unchen, Germany}
\affiliation{$^{10}$Max Planck Institute for Astronomy, Königstuhl 17, 69117 Heidelberg, Germany}
\affiliation{$^{11}$National Radio Astronomy Observatory, Pete V. Domenici Array Science Center, Socorro, NM 87801, USA}
\affiliation{$^{12}$Centro de Astrobiolog\'{\i}a (CAB), CSIC-INTA, Ctra. de Ajalvir km 4, Torrej\'on de Ardoz, E-28850, Madrid, Spain}
\affiliation{$^{13}$Canadian Institute for Theoretical Astrophysics, University of Toronto, 60 St George St, Toronto, ON M5S 3H8, Canada}
\affiliation{$^{14}$Department of Physics, 60 St George St, University of Toronto, Toronto, ON M5S 3H8, Canada}
\affiliation{$^{15}$Dipartimento di Fisica e Astronomia - Università degli Studi di Firenze, Via Giovanni Sansone, 1, 50019 Sesto Fiorentino, Italy}
\affiliation{$^{16}$MIT Kavli Institute for Astrophysics and Space Research, 70 Vassar Street, Cambridge, MA 02139, USA}
\affiliation{$^{17}$DTU Space, Technical University of Denmark, Elektrovej 327, DK-2800 Kgs. Lyngby, Denmark}
\affiliation{$^{18}$Institute of Science and Technology Austria (ISTA), Am Campus 1, 3400 Klosterneuburg, Austria}
\affiliation{$^{19}$Geneva Observatory, University of Geneva, Chemin Pegasi 51, 1290 Versoix, Switzerland}
\affiliation{$^{20}$Waseda Institute for Advanced Study (WIAS), Waseda University, 1-21-1, Nishi-Waseda, Shinjuku, Tokyo 169-0051, Japan}
\affiliation{$^{21}$Kavli Institute for the Physics and Mathematics of the Universe (WPI),The University of Tokyo Institutes for Advanced Study, The University of Tokyo, Kashiwa, Chiba 277-8583, Japan}
\affiliation{$^{22}$Department of Physics, University of California Santa Barbara, Santa Barbara, CA 93106, USA}
\affiliation{$^{23}$Centre for Astrophysics and Supercomputing, Swinburne University of Technology, Melbourne, VIC 3122, Australia}
\affiliation{$^{24}$Kapteyn Astronomical Institute, University of Groningen, 9747 AD Groningen, The Netherlands}
\affiliation{$^{25}$Department of Physics and Astronomy, Tufts University, 574 Boston Avenue, Suite 304, Medford, MA 02155, USA}
\affiliation{$^{26}$INAF-Osservatorio Astronomico di Roma, via di Frascati 33, I-00040, Monteporzio Catone, Italy}
\affiliation{$^{27}$Dipartimento di Fisica, Universit\'a di Roma La Sapienza P.le Aldo Moro 2, I-00185 Roma, Italy}
\affiliation{$^{28}$Department of Physics and Astronomy, University of Missouri, 701 S. College Ave., Columbia, MO 65203}
\affiliation{$^{29}$ Institute for Astronomy, University of Edinburgh, Royal Observatory, Blackford Hill, Edinburgh EH9 3HJ, UK}
\affiliation{$^{30}$ Steward Observatory, University of Arizona, 933 N. Cherry Avenue, Tucson, AZ 85721, USA}

\footnotetext[*]{Corresponding author. Email: seiji.fujimoto@utoronto.ca}

\begin{abstract}
We report a long-lived nuclear transient in GNz7q, a red quasar at $z=7.19$ powered by a $\sim3\times10^{7}\,M_\odot$ black hole and hosted by a compact, dusty starburst galaxy.
The transient was identified in more than two decades of multi-epoch imaging with \textit{HST}, \textit{Spitzer}, and \textit{JWST}.
After a rapid rise, the source faded smoothly by $\Delta m\simeq1.1$ mag in the rest-frame ultraviolet over a rest-frame timescale of $\sim2$ yr, with coherent, wavelength-dependent evolution across 1--5 $\mu$m that places GNz7q above the 99.9th percentile of the SDSS quasar variability distribution and is absent at $z\gtrsim5$.
The duration and energetics disfavor superluminous supernovae, and Monte Carlo tests show that stochastic quasar variability reproduces the observed multi-band evolution with a probability of $\lesssim10^{-5}$.
A panchromatic model combining a thermal continuum with a $t^{-5/3}$ decline reproduces the light curves, yielding a peak temperature of $\simeq1.4\times10^{4}$ K, a peak bolometric luminosity of $\simeq3\times10^{45}$ erg s$^{-1}$, and a radiated energy of $\simeq1.5\times10^{53}$ erg, which imply the tidal disruption of a star of a few solar masses and place the event among the most energetic tidal disruption events (TDEs) known.
A redshifted, extremely broad Balmer-line component in independent \textit{JWST}/NIRSpec spectroscopy is consistent with a transient, non-virialized broad-line region.
\textit{JWST}/MIRI photometry further reveals delayed mid-infrared emission from $\simeq1500$ K dust at a sub-parsec radius, consistent with a dust echo of the flare and providing a direct constraint on circumnuclear dust reprocessing in a quasar at $z>7$.
The relatively low black-hole mass ($M_{\rm BH}\lesssim10^{8}\,M_\odot$) and dense star-forming nucleus of GNz7q are conditions under which TDEs are expected to be most efficient, extending the connection between energetic nuclear transients and dusty star-forming hosts to the epoch of reionization.
These observations provide a time-domain view of episodic black-hole fueling and its dusty nuclear environment 700 million years after the Big Bang.
Wide-field time-domain surveys with \textit{Roman} and \textit{Euclid} in the near-infrared, complemented by LSST at lower redshifts, may uncover such transients in large numbers and enable systematic time-domain studies of black-hole growth in the reionization era.
\end{abstract}

\section{Introduction}
\label{sec:intro}

Accreting black holes grow through episodic mass inflow, making time-domain observations a direct probe of black-hole assembly.
Supernovae trace the deaths of early stellar generations and the chemical enrichment that follows.
Tidal disruption events (TDEs) probe otherwise dormant or quiescent massive black holes and their nuclear environments \citep{gezari2021}.
Variability in active galactic nuclei (AGN) follows the accretion flow itself and the timescales on which the nuclear fuel supply changes.
At $z>6$, where supermassive black holes are already in place \citep[e.g.,][]{banados2018,wang2021}, such studies are further hindered by observational challenges, as the long rest-frame timescales of interest require decade-long monitoring in the observed frame.
As a result, the time-domain behavior of black-hole growth during the epoch of reionization remains largely unexplored.

Large-amplitude variability associated with transient accretion events has been observed in quasars at $z\lesssim2$ \citep[e.g.,][]{macleod2019,gezari2021}, but comparable phenomena have not been established at earlier cosmic epochs.
\textit{JWST} has begun to change this picture.
The JADES Transient Survey identified 79 supernovae in the GOODS-S deep field, among them a spectroscopically confirmed Type~IIP supernova at $z=3.61$ \citep{decoursey2025}.
Deep \textit{JWST} spectroscopy has also uncovered an abundant population of faint broad-line AGN at $z=4$--7, including the little red dots \citep[e.g.,][]{harikane2023agn,matthee2024,greene2024,maiolino2024jades, fujimoto2024uncover}.
Constraints on the variability of these early black holes are still very limited.
A systematic search of \textit{WISE} light curves for the 531 known quasars at $z>5.3$ recovered only rare, low-amplitude continuum variability below about $0.2$ mag and no large-amplitude events \citep{leung2026}.
Repeat \textit{JWST} spectroscopy of little red dots at $z\sim7$ has measured broad-line equivalent-width changes of about 20 percent over 2.4 rest-frame years \citep{furtak2025}, and continuum and broad-line changes of about 30 percent over roughly 13 rest-frame days \citep{lambrides2026}.
A candidate TDE at a photometric redshift of $z\simeq5$ has recently been reported in the COSMOS-Web survey \citep{karmen2025}, but sparse temporal sampling and the lack of spectroscopic confirmation leave its redshift and classification uncertain.
Spectroscopically confirmed TDEs remain limited to $z\lesssim1.2$ \citep{andreoni2022,zhu2026}, and no large-amplitude nuclear transient has been established in a quasar in the reionization era.

We identified the event reported here by chance.
While comparing new \textit{JWST}/NIRCam imaging of the GOODS-North field with archival \textit{HST} images, we found that GNz7q, a quasar at $z=7.19$ hosted by a dusty star-forming galaxy \citep{fujimoto2022,fei2026}, had become about one magnitude fainter.
To place that fading in a longer temporal context, we assembled two decades of archival multi-epoch photometry from \textit{HST}, \textit{Spitzer}, and Subaru, together with new \textit{JWST}/NIRCam and MIRI imaging and repeat \textit{JWST}/NIRSpec spectroscopy.
This paper reports a long-lived nuclear transient in GNz7q whose evolution supports a TDE interpretation, together with delayed mid-infrared emission consistent with a dust echo of the flare.

This paper is structured as follows. Section~\ref{sec:obs} describes the target and the multi-epoch observations. Section~\ref{sec:transient} presents the discovery and characterization of the long-lived nuclear transient, and Section~\ref{sec:tde} presents the light-curve modeling, two-epoch spectroscopy, dust-echo analysis, and energetics that support a TDE interpretation. We discuss the implications in Section~\ref{sec:discussion}. Tests of photometric systematics and of alternative scenarios are presented in Appendices~\ref{sec:test} and \ref{sec:alternative}.

In this paper, we adopt cosmological parameters measured by the Planck mission \citep{planck2014}, i.e.\ a $\Lambda$ cold dark matter ($\Lambda$CDM) model with total matter, 
vacuum and baryonic densities in units of the critical density,
$\Omega_{\Lambda}=$ 0.692,
$\Omega_{\rm m}$ =  0.308, 
$\Omega_{\rm b} =$ 0.0481, 
and Hubble constant, $H_{0}=100$ $h$\,km\,$s^{-1}$\,Mpc$^{-1}$, 
with $h= 0.678$. 
Based on these parameters, we adopt the angular size distance of 5.26 kpc/arcsec at the source redshift of $z=7.1899$ in this paper. 

\section{Target and Observations}
\label{sec:obs}

\subsection{The quasar GNz7q}
\label{sec:target}

\begin{figure*}
\begin{center}
\includegraphics[angle=0,width=1.0\textwidth]{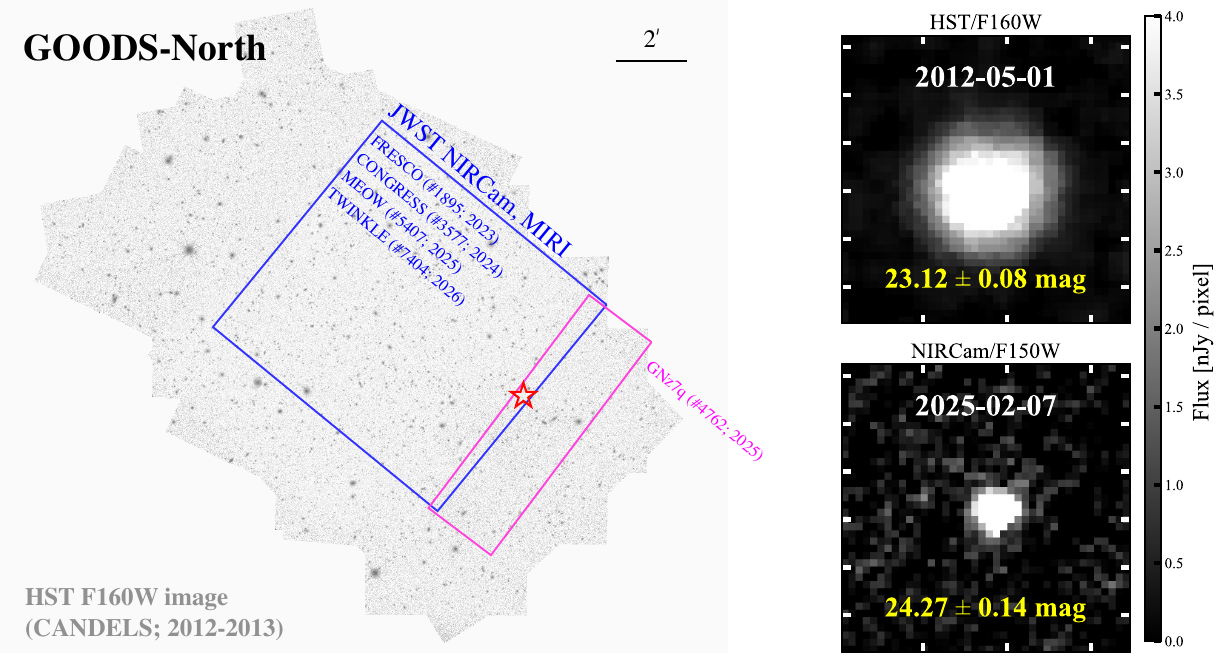}
\end{center}
\vspace{-0.2cm}
\caption{\small 
\textbf{Long-term near-infrared variability of GNz7q revealed by \textit{JWST}/NIRCam and archival \textit{HST} imaging.}
\textbf{Left:} Footprints of recent \textit{JWST} observations (2023--2026) overlaid on the \textit{HST}/F160W image from 2012--2013 in the GOODS-North field. Survey names, program IDs, and observing years are labeled, with additional details summarized in Table~\ref{tab:obs_summary}. The red star marks the position of the low-luminosity red quasar \targ\ at $z=7.19$ \citep{fujimoto2022,fei2026}.
\textbf{Right:} Multi-epoch imaging demonstrates a pronounced decade-long dimming of \targ. The top and bottom panels show \textit{HST}/F160W imaging from 2012 and \textit{JWST}/NIRCam F150W imaging from 2025, respectively, highlighting the significant flux decrease observed in \targ\ between the two epochs. Both panels display $2''\times2''$ cutouts.
}
\label{fig:nircam_entire}
\end{figure*}

The compact, dusty object GNz7q was initially identified as a low-luminosity, red quasar candidate in deep \textit{Hubble Space Telescope} (\textit{HST}) imaging of the Great Observatories Origins Deep Survey (GOODS) North field through a uniform reprocessing of all archival datasets and 1-mm follow-up spectroscopy with NOEMA \citep{fujimoto2022}. Figure~\ref{fig:nircam_entire} marks its position in the \textit{HST}/F160W image. 
Intensive spectroscopic and multi-wavelength follow-up observations were carried out between January and May 2025 to investigate the nature of GNz7q (see Sections~\ref{sec:nircam} and \ref{sec:miri}). 
These observations confirm \targ\ to be a unique red quasar: moderately luminous in the rest-frame UV (absolute magnitude $M_{1450}=-22.7$) yet extremely faint in X-rays (optical-to-X-ray spectral index $\alpha_{\rm ox}<-2.23$ at the 3$\sigma$ level), embedded in a dusty starburst galaxy with a star-formation rate of ${\rm SFR}=330\pm100~M_\odot\,\mathrm{yr^{-1}}$ and a stellar mass of $\log(M_{\rm star}/M_{\odot}) = 10.5\pm0.4$ at $z=7.1899$ \citep{fujimoto2022,fei2026}.
Prominent broad Balmer emission lines are detected, yielding the BH mass of $\log(M_{\rm BH}/M_{\odot}) = 7.5\pm0.3$ based on multiple empirical virial calibrations. %
A decomposition of the composite spectral energy distribution (SED) into AGN and host-galaxy components yields an AGN bolometric luminosity of $\log(L_{\rm bol}) = 46.2 \pm 0.1~{\rm erg~s^{-1}}$ and a super-Eddington ratio of $\lambda_{\rm Edd} = 2.7 \pm 0.4$. 
These properties indicate that \targ\ is currently undergoing a high accretion phase, representing an early stage in quasar evolution. 
Given its extreme accretion rate and intense star-formation activity, GNz7q is expected to evolve into a luminous quasar system hosting a supermassive black hole ($>10^{9}\,M_{\odot}$) embedded in a massive galaxy at later epochs \citep{fei2026}.

\subsection{\jwst/NIRCam observations}
\label{sec:nircam}

GNz7q was observed with \jwst/NIRCam as part of three independent General Observer (GO) programs: three blind NIRCam grism surveys of the GOODS-North field conducted in Cycle~1 (GO-1895; PI: P.~Oesch), Cycle~2 (GO-3577; PIs: E.~Egami \& F.~Sun), and Cycle~4 (GO-7404; PIs: R.~Naidu, J.~Matthee, \& J.~Chisholm), and a targeted Cycle~3 follow-up program (GO-4762; PIs: S.~Fujimoto \& G.~Brammer). 
The Cycle~1 GO-1895 observations, obtained on 2023 February 12, and the Cycle~2 GO-3577 observations, obtained on 2024 February 12, were designed as wide-area NIRCam grism mosaics of the GOODS-North field, within whose footprints GNz7q was fortunately included. 
The Cycle~4 GO-7404 observations were later carried out on 2026 January 28 as part of multi-epoch monitoring of the GOODS-North field using the same footprint as GO-1895.
The Cycle~3 observations (GO-4762), obtained on 2025 February 7, were executed as dedicated follow-up observations of GNz7q motivated by its unique quasar and host-galaxy properties \citep{fujimoto2022}.  
In all three programs, the NIRCam data were acquired in conjunction with NIRCam grism spectroscopy. 
Imaging data were obtained either as direct images for astrometric and wavelength calibration or as paired short-wavelength (SW) exposures taken simultaneously with the long-wavelength (LW) grism observations, providing contemporaneous imaging during the spectroscopic integrations.

In the blind GOODS-North surveys (GO-1895, GO-3577, and GO-7404), NIRCam grism observations were carried out using the LW F444W, F410M, or F356W filters in combination with SW filters including F210M, F182M, F115W, and F090W. 
Each grism exposure employed the GRISMR configuration with SHALLOW4 or MEDIUM2 readout patterns, 7--9 groups per integration, one integration per exposure, and four primary dithers. 
Direct imaging exposures were obtained with the F444W or F356W filter to provide astrometric and wavelength calibration. 
The total on-source exposure times for the grism observations in these programs were $\simeq1.5$--$3.5$~ks per filter configuration.
The Cycle~3 follow-up program (GO-4762) adopted a similar grism observing strategy but with a filter set optimized for continuum and emission-line characterization of GNz7q. 
Grism observations were obtained using the LW F410M filter paired with the SW F210M and F182M filters, again using the GRISMR configuration, a MEDIUM2 readout pattern, 7 groups per integration, one integration per exposure, and four primary dithers. 
In addition, short direct imaging exposures were acquired using the F356W/F150W and F444W/F182M filter pairs on the SW and LW channels, respectively, with SHALLOW2 or SHALLOW4 readout patterns. 
The total grism exposure time in the Cycle~3 observations was $\simeq2.7$~ks per filter configuration. 
A summary of all NIRCam photometric observations used in this work is provided in Table~\ref{tab:obs_summary}.

The accompanying NIRCam imaging data were reduced using a consistent procedure for both programs, starting from the uncalibrated files processed with the \jwst\ Science Calibration Pipeline (v12.0.9) and the CRDS context \texttt{jwst\_1414.pmap}. 
In addition to the standard STScI pipeline steps, we applied custom corrections to mitigate instrumental artifacts commonly affecting NIRCam imaging, including the removal of snowballs, wisps, residual cosmic rays, persistence, and diffraction spikes, together with an initial correction for $1/f$ noise. 
After Stage~2 processing, a secondary $1/f$ noise correction was performed using sigma-clipping, followed by two-dimensional background subtraction implemented with the \texttt{sep} package. 
Astrometrically aligned mosaics were then produced through Stage~3 processing, referenced to \textit{Gaia} stars, with final pixel scales of 20~mas~pix$^{-1}$ for the short-wavelength channel and 40~mas~pix$^{-1}$ for the long-wavelength channel.
For the purpose of the time-domain analyses in this study, the imaging data from GO-1895 and GO-4762 observations were mosaicked independently, yielding separate epoch-specific images.

Aperture photometry was performed on the NIRCam mosaics to measure the flux of \targ\ at each wavelength and epoch. 
\targ\ appears as an isolated, unresolved point source in all NIRCam bands, allowing for straightforward aperture-based measurements. 
We measured the source fluxes within circular apertures of radius $r=0\farcs4$, and estimated local backgrounds using surrounding annuli free of contaminating sources. 
To recover total fluxes, we applied aperture corrections derived from the NIRCam point-spread function (PSF), using encircled-energy curves appropriate for each filter and detector channel\footnote{\url{https://jwst-docs.stsci.edu/jwst-near-infrared-camera/nircam-performance/nircam-point-spread-functions\#gsc.tab=0}}. 
The same photometric procedure and aperture corrections were applied consistently across all epochs and filters to ensure robust relative flux comparisons. 
Photometric uncertainties include contributions from background noise, Poisson statistics, and uncertainties in the aperture corrections, and are propagated throughout the subsequent variability analysis.

\subsection{\jwst/MIRI observations}
\label{sec:miri}

Mid-infrared imaging of GNz7q was obtained with \jwst/MIRI as part of two independent GO programs. 
The first set of observations was carried out in the Cycle~3 program GO-4762 (PIs: S.~Fujimoto \& G.~Brammer) on 2025 February 19, targeting GNz7q with the F1280W filter as a dedicated follow-up for \targ\ \citep{fei2026}. 
In addition, GNz7q was observed within a wider MIRI mosaicking program, GO-5407 (PIs: G.~Leung, R.~Endsley, \& S.~Finkelstein), on 2025 February 22, which obtained imaging in the F1000W and F2100W filters, providing complementary mid-infrared wavelength coverage at a similar epoch \citep{leung2026meow}.
All MIRI observations were performed in full-frame imaging mode using cycling dither patterns to mitigate detector artifacts and improve spatial sampling. 
The F1280W observations employed a FASTR1 readout pattern with multiple integrations per exposure and six primary dithers, resulting in a total on-source exposure time of $\simeq4.1$~ks. 
The F1000W and F2100W observations were obtained using FASTR1 readout patterns with four primary dithers, with total exposure times of $\simeq0.7$~ks and $\simeq3.1$~ks, respectively. 
A summary of the MIRI observations is provided in Table~\ref{tab:obs_summary}.

\targ\ is unresolved and isolated also in the MIRI images, allowing for straightforward aperture-based measurements. 
Fluxes were measured within circular apertures of radius $r=1\farcs0$, with local backgrounds estimated from surrounding source-free annuli. 
To recover total fluxes, aperture corrections appropriate for each filter (F1000W, F1280W, and F2100W) were applied using the encircled-energy curves provided in the latest CRDS reference file\footnote{jwst\_miri\_apcorr\_0014.fits}. 
The resulting aperture-corrected fluxes form the basis of the dust-echo analysis presented in this work.

\subsection{Ancillary data (Subaru, \textit{Spitzer}, and \textit{HST})}
\label{sec:ancillary}

In addition to the new \jwst\ observations, we assembled all relevant archival imaging data of \targ\ from Subaru, \textit{HST}, and \textit{Spitzer} that are publicly available and cover the source position. These ancillary datasets provide critical temporal baselines spanning more than two decades in the observed frame.

\paragraph{Subaru/MOIRCS:} 
Near-infrared $JHK_\mathrm{s}$ imaging from the MOIRCS Deep Survey (MODS; \citealt{kajisawa2011}) was available, in which \targ\ falls within the survey mosaic and is detected in the publicly released source catalog. Because these data effectively constitute a single epoch for the purposes of the present time-domain analysis, we adopted the total flux measurements directly from the published MODS catalog, without reprocessing the imaging data. The corresponding fluxes are listed in Table~\ref{tab:obs_summary}.

\paragraph{\textit{HST}/WFC3:} 
We collected all archival \textit{HST}/WFC3 imaging observations in which \targ\ falls within the field of view, focusing on the F125W and F160W filters whose wavelength ranges are close to the NIRCam/F115W and F150W filters and probe the rest-frame ultraviolet continuum at $z\simeq7.2$. Individual calibrated exposures were retrieved from the \textit{HST} archive and processed in a uniform manner using the \texttt{grizli} pipeline, following the same procedures adopted in \citet{fujimoto2022}. In brief, the images were astrometrically aligned to a common reference frame and drizzled onto a consistent pixel grid, with cosmic-ray rejection and background subtraction performed at the exposure level.
To enable a time-domain analysis, the reduced exposures were grouped into distinct epochs based on their observation dates, with data separated by more than several months treated as independent epochs. For each epoch, the aligned exposures were combined into an integrated mosaic using inverse-variance weighting. Aperture photometry was then performed on the epoch-level mosaics using a circular aperture of radius $r = 0\farcs4$, appropriate for an unresolved source. Because \targ\ is consistent with a point source in all \textit{HST} bands, aperture corrections were derived from the encircled-energy curves appropriate for each filter\footnote{\url{https://www.stsci.edu/hst/instrumentation/wfc3/data-analysis/photometric-calibration/ir-encircled-energy}}. 
The resulting epoch-resolved photometry is reported in Table~\ref{tab:obs_summary}.

\paragraph{\textit{Spitzer}/IRAC \& MIPS:} 
We carried out an analogous analysis for archival \textit{Spitzer}/IRAC channel~1 (3.6~$\mu$m) and channel~2 (4.5~$\mu$m) imaging. All public IRAC observations covering the position of \targ\ were identified, including the GOODS \citep{dickinson2004}, SEDS \citep{ashby2013}, S-CANDELS \citep{ashby2015}, and GREATS \citep{stefanon2021} programs. At the exposure level, the data were processed uniformly using the \texttt{grizli}/\texttt{golfir} framework, following the same methodology as described in \citet{fujimoto2022}, including astrometric alignment to the \textit{HST} reference frame and background modeling.
For the time-domain analysis, the IRAC data were grouped into epochs based on their observed-frame dates, with a characteristic grouping timescale of approximately one year. Within each epoch, individual Astronomical Observation Request (AOR) mosaics were combined into a single stacked image using inverse-variance weighting. The effective epoch time was defined as the exposure-time-weighted mean observation date of the contributing AORs.
Photometry was measured on the epoch-level IRAC mosaics using a circular aperture of radius $r = 2\farcs4$. Given the compact, unresolved nature of \targ\ in the IRAC images, standard aperture corrections were applied to convert the measured aperture fluxes into total fluxes. 
For MIPS 24~$\mu$m, the data effectively constitute a single epoch, similar to Subaru/MOIRCS data, and we employed the total flux measurements directly from the value presented in \citet{fujimoto2022}, without reprocessing the imaging data.
The resulting multi-band and multi-epoch IRAC photometry is summarized in Table~\ref{tab:obs_summary}.

\subsection{\jwst/NIRSpec spectroscopy}
\label{sec:nirspec}

GNz7q was observed with \jwst/NIRSpec in the multi-object spectroscopy mode as part of GO-4762 on 2025 May 17, using the G140M/F070LP and G395M/F290LP gratings with on-source exposure times of 7.4~ks per grating.
These observations, their reduction, and the emission-line properties of GNz7q are presented in \citet{fei2026}.
GNz7q was subsequently re-observed with the G395M grating by the independent program GO-9214 (PIs: C.~Mason \& D.~Stark) on 2026 May 12, 44 rest-frame days after the GO-4762 epoch.
We reduced the GO-9214 data in the same manner as in \citet{fei2026} and use the two epochs to test for spectral variability of the broad Balmer emission (Section~\ref{sec:rblr}).

\setlength{\tabcolsep}{3pt}
\begin{table*}
\centering
\caption{Summary of available photometric data of \targ\ used in this study\label{tab:obs_summary}}
{%
\renewcommand{\arraystretch}{0.9}
\begin{tabular}{lccccc}
\hline
Filter & Program ID (Survey) & Observation Date & Exposure & Flux & Ref.$^{\dagger}$ \\
       &                  &  [UT]            &     [k~sec]  &   [$\mu$Jy] & \\ \hline
\multicolumn{6}{c}{\bf Subaru/MOIRCS} \\ 
J & S07A-010 (MODS) & 2006-04$\sim$05 & 28.8 & $1.09 \pm 0.06$ & K11 \\
H & S07A-010 (MODS) & 2006-04$\sim$05 & 9.0  & $1.43 \pm 0.11$ & K11 \\
K$_\mathrm{s}$ & S07A-010 (MODS) & 2006-04$\sim$05 & 29.9 & $2.78 \pm 0.04$ & K11 \\ \hline
\multicolumn{6}{c}{\bf \textit{HST}/WFC3} \\ 
F125W (epoch1) & 12443 (CANDELS)  & 2012-04$\sim$05 & 2.1 & $1.23 \pm 0.12$ & G11, This \\
F125W (epoch2) & 12444 (CANDELS)  & 2012-11-06 & 1.2 & $1.15\pm0.12$ & G11, This \\
F125W (epoch3) & 12445 (CANDELS)  & 2013-08-06 & 1.2  & $1.08 \pm 0.12$ & G11, This \\
F160W (epoch1) & 12443 (CANDELS)  & 2012-04$\sim$05 & 3.0 & $2.06 \pm 0.16$ & G11, This \\
F160W (epoch2) & 12444 (CANDELS)  & 2012-11-06 & 1.2  & $1.95 \pm 0.16$ & G11, This \\
F160W (epoch3) & 12445 (CANDELS)  & 2013-08-06 & 0.6 & $1.91 \pm 0.16$    & G11, This  \\ \hline
\multicolumn{6}{c}{\bf \textit{Spitzer}/IRAC} \\ 
Ch1 3.6$\mu$m (epoch1) & 169 (GOODS)     & 2004--2005  & 297.3 & $3.05 \pm 0.06$    & D03, This \\
Ch1 3.6$\mu$m (epoch2) & 61040 (SEDS)      & 2010--2011 & 49.1 & $2.54 \pm 0.21$    & A13, This \\
Ch1 3.6$\mu$m (epoch3) & 80215 (S-CANDELS) & 2012 & 102.4 & $3.89 \pm 0.11$ & A15, This \\
Ch1 3.6$\mu$m (epoch4) & 11134 (GREATS) & 2015--2016  & 481.7 & $3.81 \pm 0.08$    & S21, This \\
Ch2 4.5$\mu$m (epoch1) & 169 (GOODS) & 2004--2005 & 249.0 & $2.94 \pm 0.10$    & D03, This \\
Ch2 4.5$\mu$m (epoch2) & 61040 (SEDS) & 2010--2011 & 86.4 & $1.69 \pm 0.19$    & A13, This \\
Ch2 4.5$\mu$m (epoch3) & 80215 (S-CANDELS) & 2012 & 217.9 & $3.56 \pm 0.08$ & A15, This \\
Ch2 4.5$\mu$m (epoch4) & 11134 (GREATS) & 2015--2016  & 567.0 & $3.73 \pm 0.09$    & S21, This \\ \hline
\multicolumn{6}{c}{\bf \textit{Spitzer}/MIPS} \\ 
24$\mu$m & 169 (GOODS) & 2004-05-27 & 37.1 & $28.10 \pm 6.60$    & M11, F22  \\ \hline
\multicolumn{6}{c}{\bf \textit{JWST}/NIRCam} \\ 
F115W          & 3577 (CONGRESS) & 2024-02-12 & 2.2 & $0.42 \pm 0.07$   & E23, This \\ 
F150W          & 4762 (GNz7q) & 2025-02-07 & 0.2 & $0.71 \pm 0.09$   & F26, This \\ 
F182M (epoch1) & 1895 (FRESCO) & 2023-02-12 & 3.8 & $1.37\pm0.12$ & O23, This \\
F182M (epoch2) & 4762 (GNz7q) & 2025-02-07 & 3.0 & $1.20\pm0.12$ & F26, This \\ 
F182M (epoch3) & 7404 (TWINKLE) & 2026-01-28 & 1.8 & $1.14\pm0.11$ & N25, This \\ 
F210M (epoch1) & 1895 (FRESCO) & 2023-02-12  & 3.5 & $1.87 \pm 0.15$ & O23, This \\ 
F210M (epoch2) & 4762 (GNz7q) & 2025-02-07 & 2.7 & $1.65 \pm 0.14$ & F26, This \\ 
F210M (epoch3) & 7404 (TWINKLE) & 2026-01-28 & 1.5 & $1.53 \pm 0.13$ & N25, This \\ 
F356W (epoch1) & 3577 (CONGRESS) & 2024-02-12 & 0.3 & $3.16 \pm 0.19$ & E23, This \\ 
F356W (epoch2) & 4762 (GNz7q) & 2025-02-07 & 0.2 & $3.02 \pm 0.19$   & F26, This \\ 
F444W (epoch1) & 1895 (FRESCO) & 2023-02-12 & 0.3 & $3.52 \pm 0.20$ & O23, This \\ 
F444W (epoch2) & 4762 (GNz7q) & 2025-02-07 & 0.3 & $3.33 \pm 0.20$   & F26, This \\ 
F444W (epoch3) & 7404 (TWINKLE) & 2026-01-28 & 0.3 & $3.26 \pm 0.20$   & N25, This \\ 
\hline
\multicolumn{6}{c}{\bf \textit{JWST}/MIRI} \\ 
F1000W & 5407 (MEOW)  & 2025-02-22 & 0.7 & $5.61 \pm 0.09$    & L26, This \\
F1280W & 4762 (GNz7q) & 2025-02-19 & 4.1 & $10.34 \pm 0.09$   & F26, This \\
F2100W & 5407 (MEOW)  & 2025-02-22 & 3.1 & $25.01 \pm 0.29$   & L26, This \\
\hline
\end{tabular}
\vspace{0.1cm}
}
\begin{minipage}{\linewidth}
\vspace{0.2cm}
{%
\raggedright
$\dagger$ References --
\citet{kajisawa2011} (K11); 
\citet{grogin2011} (G11); 
\citet{dickinson2003} (D03); 
\citet{ashby2013} (A13); 
\citet{ashby2015} (A15); 
\citet{stefanon2021} (S21); 
\citet{magnelli2011} (M11); 
\citet{fujimoto2022} (F22); 
\citet{oesch2023} (O23); 
\citet{fei2026} (F26); 
\citet{naidu2025a} (N25);
\citet{egami2023} (E23);
\citet{leung2026meow} (L26). 
``This'' denotes photometry newly measured in this work using our own systematic data reduction and processing (see Sections~\ref{sec:nircam}--\ref{sec:ancillary}). 
}
\end{minipage}
\end{table*}

\section{A Decade-long Nuclear Transient in GNz7q}
\label{sec:transient}

\begin{figure*}
\begin{center}
\includegraphics[width=0.485\textwidth]{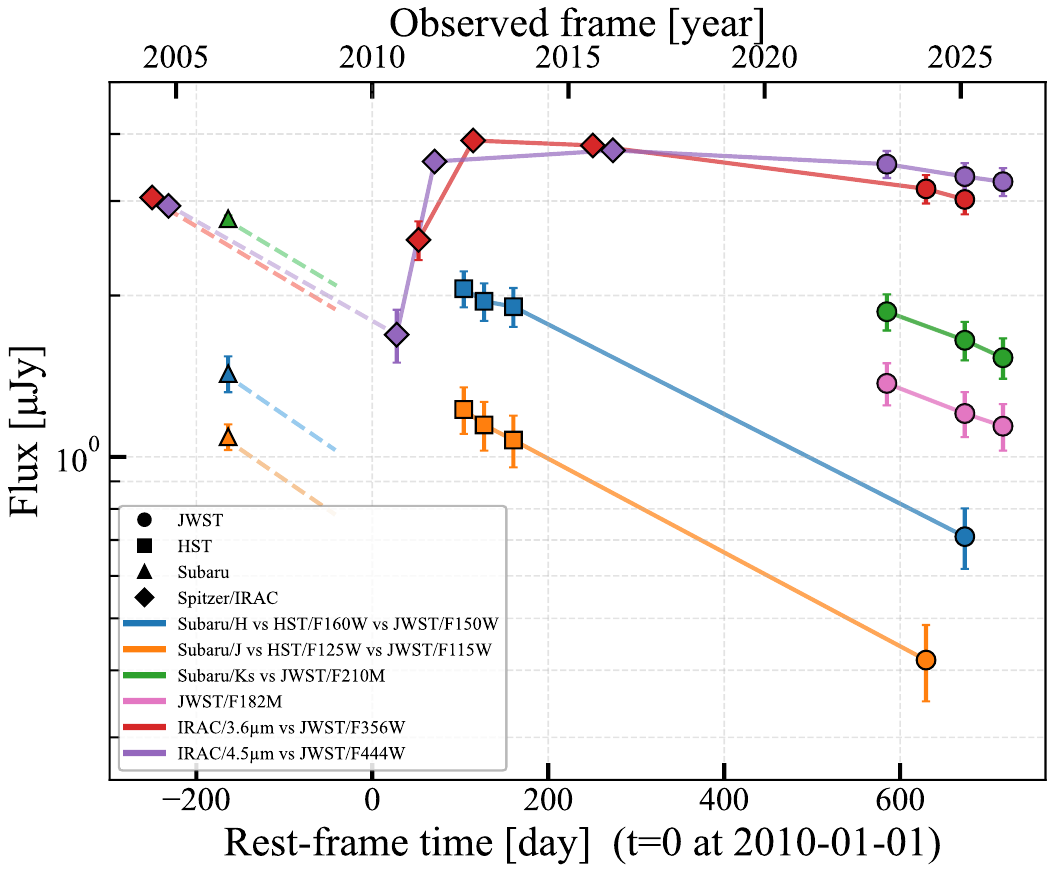}
\includegraphics[width=0.500\textwidth]{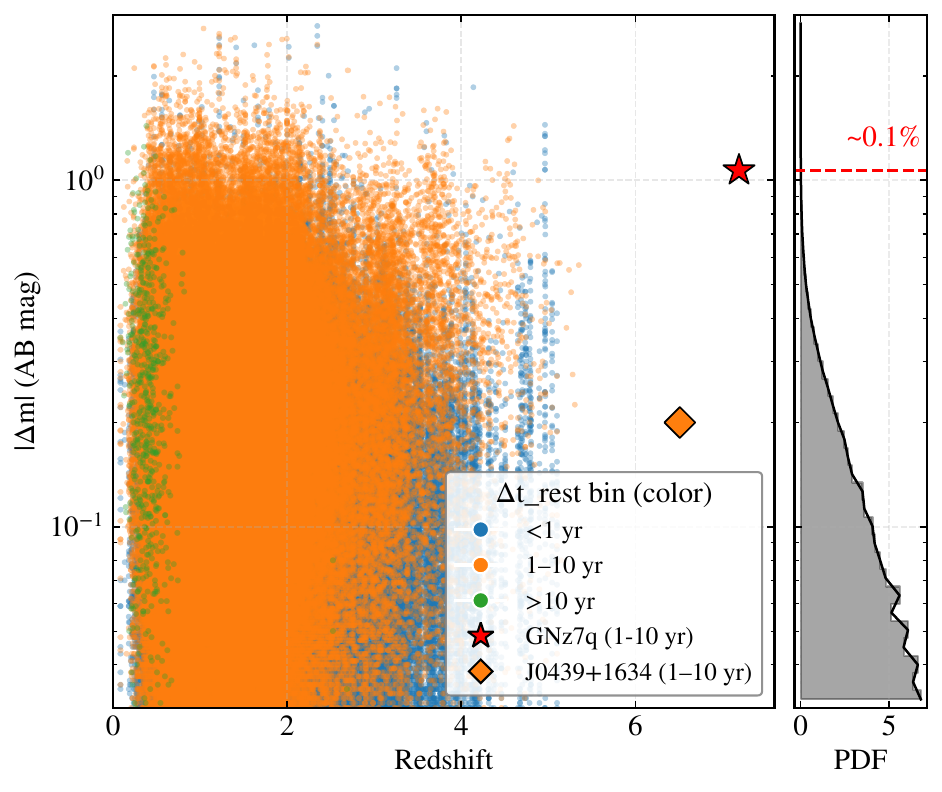}
\end{center}
\vspace{-0.2cm}
\caption{\small 
\textbf{Multi-epoch, multi-wavelength photometry over two decades, confirming the long-term, outstanding variability of GNz7q.}
\textbf{Left:} Multi-epoch photometry compiled from all available imaging, using filters closely matched in effective wavelength across different instruments (e.g., \textit{HST}/F160W versus \textit{JWST}/NIRCam F150W). 
Since around 2012, the shorter-wavelength bands exhibit a consistent, smooth dimming of up to $\Delta m\simeq1.1$\,mag over a rest-frame timescale of $\sim$2\,yr, while the fading amplitude is more modest at $\sim$4--5\,$\mu$m. 
This wavelength-dependent evolution is naturally expected for a cooling transient event. 
Data obtained prior to 2010 show additional variability and are not considered part of the main event discussed here (see text). 
Although such early variability could in principle arise from the same physical event (e.g., a multi-peaked TDE; \citealt{leloudas2016}), the sparse temporal sampling prevents a firm interpretation, and we therefore focus on the better-sampled transient evolution observed after 2012. 
Calibration, reduction, and photometry systematics are firmly ruled out based on tests with nearby sources in the same field (Figure~\ref{fig:test}).
\textbf{Right:} Comparison of the rest-UV variability amplitude with the SDSS quasar variability distribution \citep{macleod2010}, along with the probability density function (PDF) of $|\Delta m|$ collapsed over redshift. 
Points are color-coded by rest-frame baseline.  
GNz7q exhibits $>1$\,mag variability, above the 99th percentile of SDSS quasars. 
The orange diamond marks J0439+1634 at $z=6.51$, the clearest photometric variability detection among the known $z>5.3$ quasars, with an amplitude of $\simeq0.2$ mag in the rest-frame ultraviolet over $\simeq2$ rest-frame years \citep{leung2026}.
}
\label{fig:variability}
\end{figure*}

The extreme variability of GNz7q, a $z=7.19$ quasar hosted by a dusty star-forming galaxy \citep{fujimoto2022,fei2026}, was first identified through a comparison between archival \textit{HST} imaging and recent \textit{JWST}/NIRCam observations. The right panel of Figure~\ref{fig:nircam_entire} shows that, when matched at similar effective wavelengths, the source appears significantly fainter in the \textit{JWST} data than in the earlier \textit{HST} images, with declines of nearly one magnitude between \textit{HST}/F160W and NIRCam/F150W, and between \textit{HST}/F125W and NIRCam/F115W. 
A re-analysis of the full set of archival \textit{HST}/WFC3 observations reveals that this fading trend is already present within the \textit{HST} epochs alone, exhibiting a systematic, monotonic decline beginning around 2012 across multiple visits. When combined with the later \textit{JWST}/NIRCam observations, the evolution extends over more than a decade in the observed frame, reaching a total amplitude of $\Delta m\simeq1.1$\,mag by 2025 (see Figure~\ref{fig:variability}, left).

The amplitude of the observed variability places \targ\ among the most extreme nuclear transients observed in quasars. Compared with the SDSS quasar variability distribution \citep{macleod2010}, shown in the right panel of Figure~\ref{fig:variability}, \targ\ lies on the extreme tail, indicating that such large-amplitude changes occur in only $\sim$0.1\% of quasars even at lower redshift. 
At $z\lesssim2$, variability of this magnitude is associated with a range of physical mechanisms, including accretion-state changes, episodic fueling events such as TDEs, and, in some cases, variable obscuration \citep{gezari2021}. Identifying and physically interpreting such events requires multi-year monitoring in the rest frame. At high redshift, these timescales are further stretched in the observed frame, making the detection of comparable variability increasingly difficult. The variability observed in \targ\ therefore represents a rare opportunity to probe a transient accretion episode in a quasar at $z>7$.

To investigate the origin of this variability, we compiled multi-epoch photometric data spanning the rest-frame ultraviolet to mid-infrared, combining observations from \textit{HST}, \textit{Spitzer}, and \textit{JWST}, including mid-infrared coverage with \textit{JWST}/MIRI. This broad wavelength coverage enables us to trace both the evolution of the primary transient emission and its reprocessing by circumnuclear dust.

In the left panel of Figure~\ref{fig:variability}, we present multi-epoch photometry spanning more than two decades in the observed frame. Measurements obtained at similar effective wavelengths are grouped and displayed with common color coding. Despite slight differences in instrumentation and filter bandpasses, all bands spanning $\sim$1--5~$\mu$m (rest-frame UV--optical) exhibit a coherent, monotonic decline over the same decade-long baseline after 2012, with larger amplitudes at shorter wavelengths and more modest evolution at $\sim$4--5~$\mu$m. 
The consistency of this behavior across independent observatories and filters rules out calibration uncertainties or reduction artifacts. This conclusion is further supported by applying the same photometric analysis to nearby, isolated control sources, which do not show comparable variability (see Figure~\ref{fig:test}). Together, these data provide strong evidence that GNz7q underwent a genuine, coherent physical evolution.

Note that the earliest ground-based Subaru/MOIRCS $J$- and $H$-band measurements from 2006 indicate a brighter state than that observed in the most recent \textit{JWST}/NIRCam F115W and F150W data. 
This suggests additional variability prior to 2010. Such early activity could arise from supernovae in the dusty starburst host or from a multi-peaked TDE produced by the result of circularization shocks and delayed accretion \citep[e.g.,][]{leloudas2016, wli2025}. 
However, the sparse temporal sampling before 2010 prevents a definitive interpretation.

\section{Evidence for a Tidal Disruption Event and Its Dust Echo}
\label{sec:tde}

Focusing on the long-term monotonic decline observed over $\sim$2012--2025 (Figure~\ref{fig:variability}), the smooth fading over a rest-frame timescale of $\sim$2~yr, following a rapid rise, is difficult to reconcile with supernovae or stochastic AGN variability. Even the most extreme superluminous supernovae (SLSNe) typically decline on timescales of $\lesssim1$~yr in the rest frame \citep{galyam2019}, significantly shorter than observed here. 
Instead, the temporal evolution closely resembles the power-law decay expected for fallback accretion in a TDE, for which $\dot{M}\propto t^{-5/3}$ is predicted in the simplest picture \citep{rees1988}.

To further assess this scenario, we model the multi-band light curves jointly over $\sim$1--5~$\mu$m (observed frame), describing the emission as the sum of a steady AGN component and a transient component with a thermal spectrum and TDE-like temporal evolution. 

\subsection{Multi-band light-curve modeling}
\label{sec:lc_fit}

We modeled the long-term variability using all available multi-epoch photometry at $\sim$1--5~$\mu$m (observed frame), including \textit{HST}/WFC3 (F125W, F160W), \textit{Spitzer}/IRAC (3.6 and 4.5~$\mu$m), and \textit{JWST}/NIRCam (F115W, F150W, F210M, F356W, F444W).
Observation times were converted to rest-frame days relative to $t_0$, defined as the earliest observing epoch in the assembled dataset, via
$t_{\rm rest}=(t_{\rm obs}-t_0)/(1+z)$ with $z=7.19$.
To focus on the transient event showing the smooth decay, following the steep rise after around the rest-frame 200~days in Figure~\ref{fig:variability}, we restricted the fit to $t_{\rm rest}>200$~days.

We modeled the observed flux density in each band as
\begin{equation}
F_{\nu,{\rm obs}}(t) = F_{\nu,{\rm base}}({\rm band}) + F_{\nu,{\rm trans}}(t,{\rm band}),
\end{equation}
where $F_{\nu,{\rm base}}$ is a steady baseline component and $F_{\nu,{\rm trans}}$ is a transient component.
The baseline model was constructed from an AGN+host template SED \citep{fujimoto2022} ($F_{\nu,{\rm temp}}$) and forward-modeled into each band using synthetic photometry through the corresponding filter transmission curve, $F_{\nu,{\rm temp}}\rangle_{\rm band}$.
A single multiplicative normalization parameter $C_{\rm AGN}$ was fitted such that
$F_{\nu,{\rm base}}({\rm band}) = C_{\rm AGN}\times F_{\nu,{\rm temp}}({\rm band})$. 

For the transient time dependence we adopted the empirical ``Gaussian rise + power-law decay'' prescription introduced for optical TDE light curves by \citet{van-velzen2021}, which captures a rounded maximum while retaining a late-time asymptotic power-law decline.
Specifically, we modeled
\begin{multline}
F_{\nu,{\rm trans}}(t,{\rm band}) = A_{\rm ref} \times
\Phi_{\rm GP}(t_{\rm rest};t_{\rm peak},\sigma_{\rm rise},p,t_{0,{\rm pl}})\\
\times\frac{B_\nu\left(\nu_{\rm rest},T(t_{\rm rest})\right)}
{B_\nu\left(\nu_{\rm ref,rest},T(t_{\rm rest})\right)},
\end{multline}
where $A_{\rm ref}$ is the transient peak flux density (in $\mu$Jy) in a reference band (taken to be NIRCam/F150W), $\nu_{\rm rest}=(1+z)\nu_{\rm obs}$, and $B_\nu$ is the Planck function. The time profile $\Phi_{\rm GP}$ is normalized to $\Phi_{\rm GP}(t_{\rm peak})=1$ and defined as
\begin{equation}
\Phi_{\rm GP}(t_{\rm rest})=
\begin{cases}
\exp\!\left[-\dfrac{(t_{\rm rest}-t_{\rm peak})^{2}}{2\sigma_{\rm rise}^{2}}\right], & t_{\rm rest}\le t_{\rm peak},\\[4pt]
\left(\dfrac{t_{\rm rest}-t_{\rm peak}+t_{0,{\rm pl}}}{t_{0,{\rm pl}}}\right)^{p}, & t_{\rm rest}> t_{\rm peak}.
\end{cases}
\end{equation}
with $\sigma_{\rm rise}$ controlling the width of the pre-peak Gaussian rise, $t_{0,{\rm pl}}$ setting the transition timescale into the power-law regime, and $p$ the late-time decay index. We fixed $p=-5/3$ to match the canonical TDE fallback expectation \citep{rees1988,phinney1989}.
As a consistency check, allowing $p$ to vary freely when fitting only the rest-UV data points yields $p=-1.1\pm0.3$, which is consistent with $-5/3$ within $2\sigma$ and thus supports our adoption of the canonical value.
The $B_{\nu}$ term was modeled as a single-temperature blackbody with a time-evolving temperature. For the temperature evolution, rather than adopting a linear form as used in \citet{van-velzen2021}, we employed a power-law parameterization in time,
\begin{equation}
T(t_{\rm rest}) = T_0\times(t_{\rm rest}-t_{{\rm start},T})^{-\beta},
\end{equation}
clipped to a floor of $T_{\rm min}=3\times10^3$~K to avoid unphysical values.
This choice is motivated by the long temporal baseline of our data, for which a strictly linear temperature decline would continue indefinitely and can lead to unphysical behavior at late times. In contrast, a power-law evolution naturally flattens at late epochs, providing a more stable phenomenological description while remaining consistent with theoretical expectations that the effective temperature in TDE emission models can evolve gradually with time (e.g., the multi-band TDE light-curve calculations; \citealt{lodato2011}).
To reduce parameter degeneracy given the limited cadence at early times, we tied the temperature ``start'' time to the peak time via
\begin{equation}
t_{{\rm start},T} \equiv t_{\rm peak} - 200~{\rm d},
\end{equation}
so that $T_0$ corresponds to the normalization at $(t_{\rm rest}-t_{{\rm start},T})=1$day. The 200~day offset matches the adopted lower bound of the fitted time range and ensures $t_{\rm rest}-t_{{\rm start},T}$ remains well-behaved across the fitted epochs. 

\begin{figure*}
\begin{center}
\includegraphics[angle=0,width=0.75\textwidth]{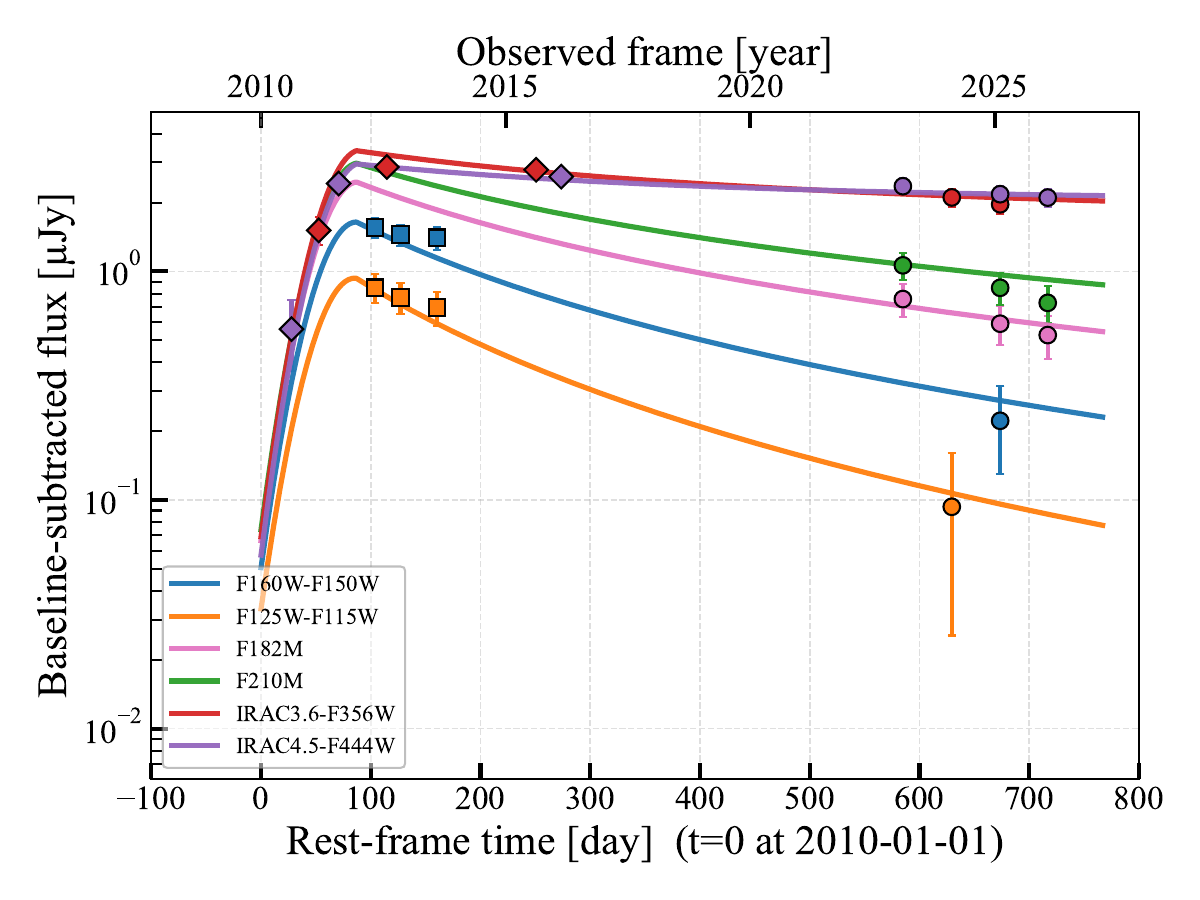}
\end{center}
\vspace{-0.2cm}
\caption{\small
\textbf{Multi-wavelength light-curve modeling of the extreme transient in GNz7q.}
Baseline-subtracted multi-epoch photometry spanning
$\sim$1--5\,$\mu$m (observed frame), combining \textit{HST}, \textit{Spitzer}, and \textit{JWST} data.
Times are shown in the rest frame.
Colored curves show the best-fit panchromatic transient model with a Gaussian rise and a $t^{-5/3}$ decay.
}
\label{fig:light_curve}
\end{figure*}

We fitted the model parameters
${T_0,\beta,A_{\rm ref},t_{\rm peak},\sigma_{\rm rise},t_{0,{\rm pl}},C_{\rm AGN}}$
by minimizing the weighted residuals between the model and the observed fluxes using a robust least-squares optimizer with a soft-$\ell_1$ loss to reduce sensitivity to outliers.
For the adopted model (Gaussian rise + power-law decay with $p=-5/3$ and power-law temperature evolution, with $t_{{\rm start},T}=t_{\rm peak}-200$~d), we obtained
$T_0=(5.7\pm0.1)\times10^4$~K, $\beta=0.262\pm0.006$, $t_{\rm peak}=87\pm6$~d, $\sigma_{\rm rise}=33\pm7$~d, $t_{0,{\rm pl}}=302\pm56$d, $A_{\rm ref}=1.64\pm0.17\,\mu$Jy, and $C_{\rm AGN}=0.30\pm0.07$, with a reduced chi-square $\chi^2_\nu=1.18$ (with $N=25$ data points and $k=7$ free parameters). In Figure~\ref{fig:light_curve}, we show that the best-fit panchromatic model reproduces the observed light curves across all bands.

We estimated the transient bolometric luminosity at peak by converting the fitted peak F150W flux density to a rest-frame specific luminosity,
\begin{equation}
L_{\nu,{\rm rest}} = \frac{4\pi D_L^2}{1+z} F_{\nu,{\rm obs}},
\end{equation}
and scaling to a bolometric luminosity assuming a blackbody using the ratio between the frequency-integrated Planck function and $B_\nu$ evaluated at the reference rest frequency.
For the best-fit parameters we obtain $T(t_{\rm peak})=(1.4\pm0.1)\times10^4$~K
and a peak bolometric luminosity $L_{\rm bol} (t_{\rm peak})=(3.5\pm0.3)\times10^{45}$ergs$^{-1}$ for the transient component.

\begin{figure}
\begin{center}
\includegraphics[angle=0,width=\columnwidth]{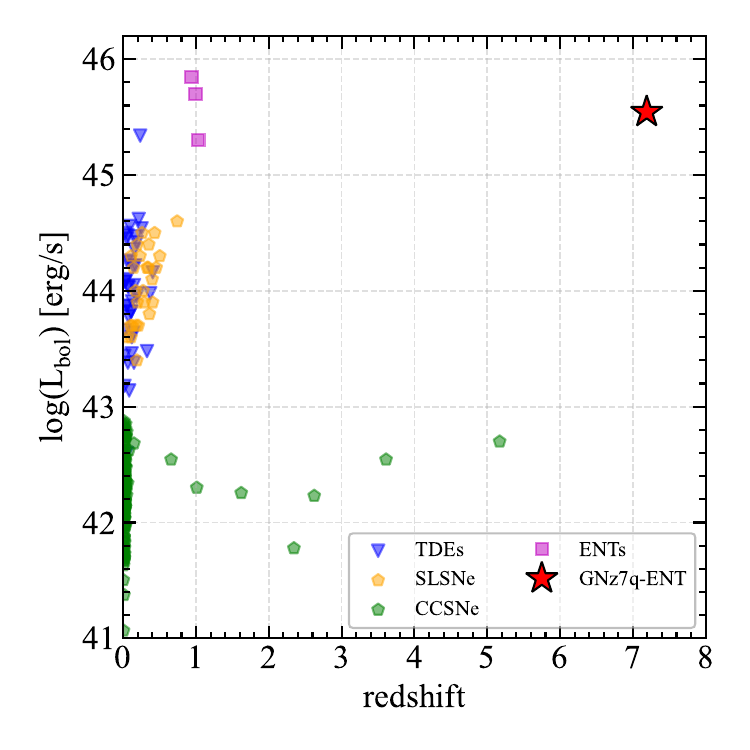}
\end{center}
\vspace{-0.2cm}
\caption{\small
\textbf{Energetics of GNz7q-ENT in the context of known transients.}
Peak bolometric luminosity as a function of redshift for a compilation of known transient events, including tidal disruption events (TDEs; \citealt{van-velzen2021,gezari2021}), superluminous supernovae (SLSNe; \citealt{decia2018}), core-collapse supernovae (CCSNe; \citealt{anderson2014,moriya2025,coulter2026}), and extreme nuclear transients (ENTs; \citealt{hinkle2025}).
}
\label{fig:Lbol_z}
\end{figure}

In Figure~\ref{fig:Lbol_z}, we compare this peak luminosity with those of known transient populations as a function of redshift. The transient is more luminous than the SLSNe in the compilation and is instead comparable to the extreme nuclear transients (ENTs) identified at $z\sim1$ \citep{hinkle2025}. We therefore refer to the event as \ent\ hereafter.

We note that no explicit dust correction has been applied in the above estimate. 
For the AGN component of \targ, a modest line-of-sight extinction of $A_V\simeq0.2$ has been inferred from the SED \citep{fei2026}. 
If we adopt the same attenuation and apply a \citet{calzetti2000} law at the rest-frame wavelength corresponding to the F150W band ($\lambda_{\rm rest}\simeq0.18\,\mu$m), the intrinsic flux increases by a factor of $\sim1.6$, leading to a dust-corrected peak luminosity of $L_{\rm bol}\simeq(5$--$6)\times10^{45}$\,erg\,s$^{-1}$. 
However, given the uncertainties in the dust geometry and the possibility that the line-of-sight extinction during the early TDE phase differs from that of the quasar continuum, we adopt the uncorrected value as our fiducial estimate and treat the dust-corrected luminosity as a reference estimate. 
Importantly, applying the dust correction places \ent\ among more extreme, luminous and energetic nuclear transients known (see Figure~\ref{fig:Lbol_z}).

\subsection{Redshifted broad Balmer component}
\label{sec:rblr}

\begin{figure}
\begin{center}
\includegraphics[angle=0,width=\columnwidth]{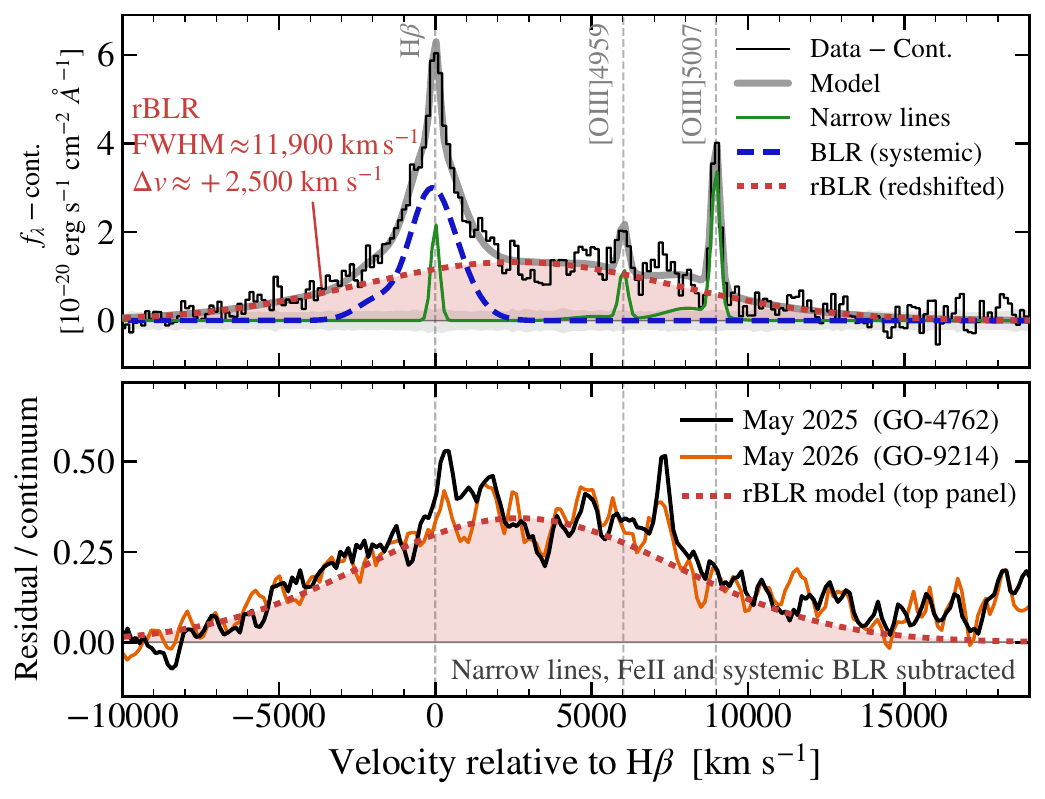}
\end{center}
\vspace{-0.2cm}
\caption{\small
\textbf{Extremely broad, redshifted Balmer component of GNz7q in two \jwst/NIRSpec epochs.}
\textbf{Top:} continuum-subtracted H$\beta$ spectrum from the deeper second epoch (GO-9214, 2026 May), decomposed into narrow lines, the systemic broad-line region (BLR) with its blueshifted absorption, and the redshifted broad component (rBLR; ${\rm FWHM}\simeq11{,}900$~km~s$^{-1}$, $\Delta v\simeq+2{,}500$~km~s$^{-1}$).
\textbf{Bottom:} the redshifted component in isolation, after removal of the narrow lines, FeII, and the systemic BLR from each epoch (GO-4762, 2025 May; GO-9214, 2026 May).
The two epochs agree in profile, width, velocity offset, and strength within the uncertainties.
}
\label{fig:rblr}
\end{figure}

Independent support is provided by the \textit{JWST}/NIRSpec spectroscopy, which reveals an additional redshifted, extremely broad Balmer-line component \citep{fei2026}.
To test whether this component varies between the two epochs described in Section~\ref{sec:nirspec}, we modeled each epoch with the same set of components as \citet{fei2026}, namely a local power-law continuum, an FeII template, the narrow lines, the systemic broad-line region with its blueshifted absorption, and the redshifted broad component. 
In the fitting, we anchored the local continuum on FeII-subtracted windows so that both epochs are referred to a common continuum.
In Figure~\ref{fig:rblr}, we show the resulting decomposition of the deeper epoch and the redshifted component isolated in each epoch.
We measure ${\rm FWHM}\simeq11{,}900$~km~s$^{-1}$ and $\Delta v\simeq+2{,}500$~km~s$^{-1}$ for this component in the deeper epoch.

The redshifted broad component is confirmed in both epochs, and the repeat spectroscopy over 44 rest-frame days shows no significant change.
The profile and equivalent width of the redshifted broad component are consistent between the two epochs within the uncertainties.
Such broad, systematically velocity-shifted emission is consistent with a transient, non-virialized broad-line region associated with TDE-driven outflows \citep{li2022}.

\subsection{Dust-echo modeling}
At $z=7.19$, the \textit{JWST}/MIRI F1000W, F1280W, and F2100W filters probe rest-frame wavelengths of $\sim1$--$3~\mu$m, corresponding to the near-peak emission of hot dust with temperatures of $\sim1000$--$3000$~K. This wavelength range is therefore well suited to trace thermal emission from circumnuclear dust heated by an intense UV--optical flare, producing a mid-infrared ``dust echo''.

Using the best-fit UV--optical transient light curve as the input radiation field, we model the mid-infrared emission with a single-temperature blackbody component (see below). Figure~\ref{fig:dust_echo} summarizes the fit to the three MIRI data points, which yields a dust temperature of $T_{\rm dust}=1485\pm32$~K and a characteristic radius of $R_{\rm d}=0.46\pm0.03$~pc. The inferred temperature is consistent with the expected sublimation temperatures of dust grains ($T_{\rm dust}\simeq1500$~K) in AGN environments \citep{barvainis1987,baskin2018}, supporting an interpretation in which the mid-infrared emission arises from thermal reprocessing of the transient.

The inferred radius is comparable to the expected scale of the inner circumnuclear dust torus, suggesting that the observed emission originates from reprocessing at sub-parsec scales. 
This provides one of the first direct constraints on circumnuclear dust reprocessing in quasars at $z>7$.
While uncertainties in dust geometry and covering factor introduce degeneracies in the inferred radius, the spectral shape across the MIRI bands tightly constrains the dust temperature. Future multi-epoch mid-infrared monitoring will enable direct measurements of the infrared lag and further constrain the structure of circumnuclear dust at $z>7$, providing a direct probe of the dust torus in quasars that might already be in place at these early epochs \citep{bosman2025}.

\label{sec:dust_echo}

To interpret the mid-infrared emission detected with \textit{JWST}/MIRI, we model the baseline-subtracted photometry as the sum of (i) direct emission from the UV--optical transient
and (ii) thermal reprocessing by circumnuclear dust, so-called \textit{dust echo}.
One may fit the infrared light curve using physically motivated
dust-echo formalisms that account for the geometry of the reprocessing region and frequency-dependent grain emissivities (e.g., the thin-shell model and modified-blackbody SED; \citealt{van-velzen2016}).
However, the present data consist of three single-epoch MIRI measurements,
so our goal is primarily a consistency check, i.e., whether the observed MIRI fluxes are compatible with dust reprocessing on plausible spatial scales, rather than a unique determination of the dust geometry.
We therefore adopt a simplified, transparent parameterization that is directly tied to the best-fit transient bolometric luminosity (Section~\ref{sec:lc_fit}) and to an analytic sublimation-radius scaling.

Although the three MIRI bands of F1000W, F1280W, and F2100W correspond to rest-frame wavelengths of $\simeq$1.2--2.6~$\mu$m at $z=7.19$, 
the Rayleigh-Jeans tail of the transient blackbody might remain non-negligible.
We therefore explicitly include the direct transient contribution,
$F_{\nu,\mathrm{trans}}(t_{\rm rest},{\rm band})$, by evaluating the best-fit
Gaussian-rise$+$power-law-decay light curve multiplied by 
the temperature-evolving single blackbody model (Section~\ref{sec:lc_fit}).

At each rest-frame time $t_{\rm rest}$, the transient bolometric luminosity
$L_{\rm bol}(t_{\rm rest})$ is inferred from the model reference-band flux density
(F150W) by assuming a single-temperature blackbody SED:
\begin{equation}
\label{eq:sed_uv}
L_{\nu,{\rm rest}}(t_{\rm rest})=\frac{4\pi D_L^2}{1+z}\,F_{\nu,{\rm obs}}(t_{\rm rest}),
\end{equation}
and scaling to the bolometric luminosity via
\begin{equation}
\label{eq:lbol}
L_{\rm bol}(t_{\rm rest})=
L_{\nu,{\rm rest}}(t_{\rm rest})
\frac{\int_0^\infty B_\nu(T)\,{\rm d}\nu}{B_\nu(\nu_{\rm ref,rest},T)}.
\end{equation}
We model the time-dependent dust-reprocessed luminosity as a convolution of the
bolometric input with a delay distribution:
\begin{equation}
\label{eq:rep}
L_{\rm bol}^{\rm rep}(t_{\rm rest})=
\int L_{\rm bol}(t_{\rm rest}-u)\,\psi(u)\,{\rm d}u,
\end{equation}
where $\psi(u)$ is a normalized top-hat response function centered at the light-travel
time delay $\tau = R_{\rm d}/c$.
We adopt a fractional width $w=f_{\rm w}\tau$ with $f_{\rm w}=0.3$ to represent modest
geometric smearing (i.e.\ a finite radial extent); $\int \psi(u)\,{\rm d}u=1$ by construction.

For the dust SED, we model the emission as a single-temperature blackbody for simplicity, given the limited spectral and temporal coverage of the current MIRI data. The monochromatic dust luminosity is 
\begin{equation}
\label{eq:sed_ir}
L_{\nu}^{\rm dust}(t_{\rm rest})=
L_{\rm bol}^{\rm rep}(t_{\rm rest})\,
\frac{B_\nu(\nu_{\rm rest},T_{\rm dust})}{\int_0^\infty B_\nu(T_{\rm dust})\,{\rm d}\nu},
\end{equation}
which is converted to observed-frame flux density through
\begin{equation}
\label{eq:flux}
F_{\nu,{\rm obs}}^{\rm dust}(t_{\rm rest})=
\frac{1+z}{4\pi D_L^2}\,L_{\nu}^{\rm dust}(t_{\rm rest}).
\end{equation}
The total model flux density is
\begin{equation}
\label{eq:f_tot}
F_{\nu,{\rm model}}(t_{\rm rest})=
F_{\nu,{\rm trans}}(t_{\rm rest}) + C_{\rm f}\times F_{\nu,{\rm dust}}(t_{\rm rest}),
\end{equation}
where $C_{\rm f}$ is a non-negative scaling factor that denotes the effective dust covering/reprocessing fraction (bounded to $0\le C_{\rm f}\le 1$ in our fiducial fits below).
In our simplified parameterization, we define $R_{\rm d}$ as the characteristic light-travel radius that sets the centroid of the
infrared response function, i.e.\ the delay $\tau = R_{\rm d}/c$
in the adopted top-hat transfer function. 
Physically, this corresponds to the effective inner radius of the dust distribution dominating the MIR emission.
We further impose a self-consistent link between this characteristic
radius and the absorbed luminosity by regarding
$R_{\rm d}$ being equal to the dust sublimation radius, $R_{\rm sub}$ and the absorbed fraction to the reprocessing amplitude, also being equal to $C_{\rm f}$, corresponding to 
\begin{equation}
\label{eq:labs}
L_{\rm abs}(t)=C_{\rm f} \times L_{\rm bol}(t). 
\end{equation}

For $R_{\rm sub}$, we adopt the analytic sublimation-radius scaling
presented by \citet{van-velzen2016}, commonly calibrated assuming a characteristic grain size of $a\simeq0.1~\mu$m:
\begin{multline}
\label{eq:rsub}
R_{\rm sub}(T_{\rm dust},L_{\rm abs}(t_{\rm peak}))=\\
R_0\left(\frac{L_{\rm abs}(t_{\rm peak})}{10^{45}\,{\rm erg\,s^{-1}}}\right)^{1/2}
\left(\frac{T_{\rm dust}}{T_0}\right)^{-\gamma},
\end{multline}
with $(R_0,T_0,\gamma)=(0.15~{\rm pc},\,1850~{\rm K},\,2.9)$.
Here $L_{\rm abs}(t_{\rm peak})$ denotes the peak bolometric luminosity absorbed by
the dust, which is equal to $C_{\rm f}\times L_{\rm bol}(t_{\rm peak})$ (Eq.~\ref{eq:labs}).

Combining Eqs.~\ref{eq:sed_uv}--\ref{eq:rsub}, the model reduces to a single
free parameter, $T_{\rm dust}$, with the dust radius and covering fraction
determined self-consistently.
Fitting the model to the three MIRI flux measurements via
Eq.~\ref{eq:f_tot}, we obtain
$T_{\rm dust}=1485\pm32$~K,
which implies
$R_{\rm d}=R_{\rm sub}=0.46\pm0.03$~pc
and
$C_{\rm f}=0.75$.
Although the present modeling adopts a simplified geometry and assumes a
single-temperature dust component, the inferred temperature is consistent
with the characteristic dust sublimation temperatures
($T_{\rm dust}\simeq1200$--$1800$~K) expected for graphite and silicate grains in AGN
environments \citep{barvainis1987,baskin2018}.
This agreement supports the interpretation that the detected MIRI emission
arises from thermal reprocessing of the central transient event observed
in the rest-frame UV--optical.

To illustrate the level of systematic uncertainty associated with relaxing the
assumption $R_{\rm d}=R_{\rm sub}$, we also fit simplified models in which
$R_{\rm d}$ is fixed to representative values (here $R_{\rm d}=0.3$~pc and
$0.5$~pc), while allowing $T_{\rm dust}$ to vary and solving for the covering
fraction $C_{\rm f}$ analytically by minimizing $\chi^2$.
These alternative models are not intended as distinct physical solutions given
the single-epoch MIR data, but rather as a visual demonstration that
comparably acceptable fits can be obtained for modest offsets in
$R_{\rm d}$. 
As shown in Figure~\ref{fig:dust_echo}, the inferred dust temperature
remains remarkably stable at $T_{\rm dust}\simeq1500$~K, even when
$R_{\rm d}$ is varied between 0.3 and 0.5~pc.
This robustness arises because the three MIRI bands probe rest-frame
$\sim1.2$--$2.6~\mu$m, which straddle the peak of a blackbody spectrum
with $T\sim1100$--$2400$~K.
Consequently, the spectral shape of the MIR emission tightly constrains
the effective dust temperature.
In contrast, $R_{\rm d}$ and $C_{\rm f}$ remain partially degenerate:
within the current data, radii in the range $\sim0.3$--$0.5$~pc can
produce similarly acceptable fits unless the additional physical
assumption $R_{\rm d}=R_{\rm sub}$ is imposed.

We emphasize that the present modeling adopts a simplified geometry
(single effective temperature and a top-hat response function).
More realistic dust configurations, including radial temperature gradients,
optical-depth effects, or alternative grain compositions, could modify the
detailed light-curve shape and dust SED, potentially shifting the inferred
values of $R_{\rm d}$ and $T_{\rm dust}$.
However, the key point illustrated in Figure~\ref{fig:dust_echo} is that
continued MIR monitoring would decisively distinguish among these scenarios
through direct measurement of the infrared lag and the time evolution of
the dust echo.
If future multi-epoch observations confirm that the dust echo follows the
 simple sublimation-radius scaling ($R_{\rm d}\propto L^{1/2}$)
expected from dust-heating equilibrium arguments \citep{barvainis1987, van-velzen2016}, such measurements could
establish dust reverberation mapping even at $z>7$ as a physically grounded probe of black-hole accretion and circumnuclear structure.
In particular, dust-reverberation measurements have been proposed
as physically grounded distance indicators for cosmological applications \citep{yoshii2014, honig2014},
offering a potential independent probe of the expansion history of the Universe.
While the present data constitute only an initial consistency check, they
demonstrate that this framework is observationally accessible even at
cosmic dawn.

\begin{figure*}
\begin{center}
\includegraphics[angle=0,width=0.92\textwidth]{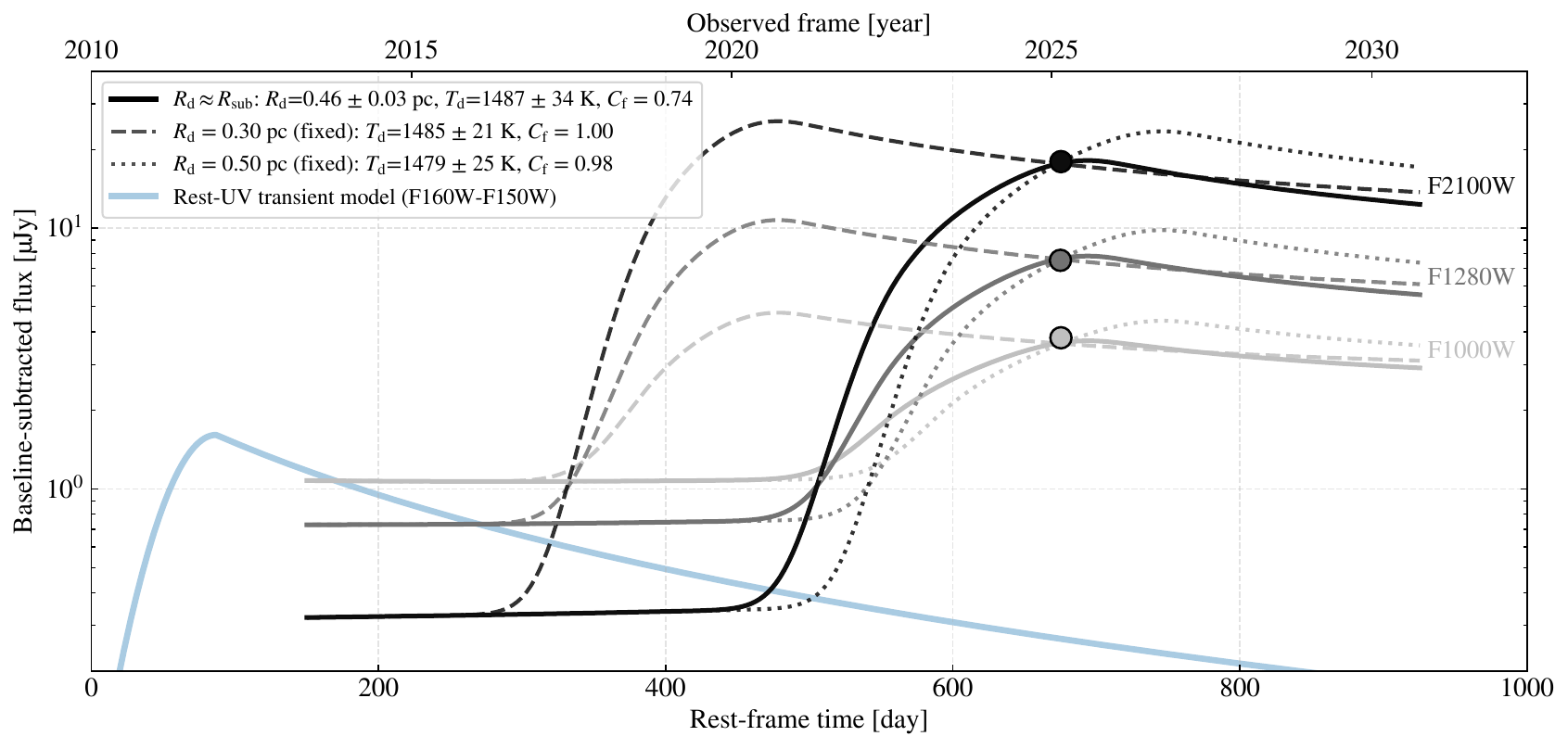}
\end{center}
\vspace{-0.2cm}
\caption{\small
\textbf{Dust-echo modeling of the \jwst/MIRI photometry.}
\jwst/MIRI F1000W, F1280W, and F2100W photometry modeled as thermal dust reprocessing of the UV--optical transient.
The solid curves show the best-fit model assuming that the characteristic dust radius equals the sublimation radius ($R_{\rm d}=R_{\rm sub}$).
Dashed and dotted curves illustrate representative solutions with fixed radii
($R_{\rm d}=0.3$ and $0.5$\,pc), highlighting the partial degeneracy arising from uncertainties in dust geometry and reprocessing efficiency.
The transparent blue curve shows the best-fit rest-frame ultraviolet transient model in the F160W--F150W band (identical to that in Figure~\ref{fig:light_curve}), illustrating the delayed mid-infrared response to the ultraviolet flare.
Importantly, future intensive mid-infrared monitoring will break the degeneracy, while the spectral shape across the current three MIRI bands already well constrains $T_{\rm dust}\simeq1500$~K.
}
\label{fig:dust_echo}
\end{figure*}

\subsection{Time evolution of the spectral energy distribution}
\label{sec:sed}

\begin{figure*}[t!]
\begin{center}
\includegraphics[angle=0,width=1.0\textwidth]{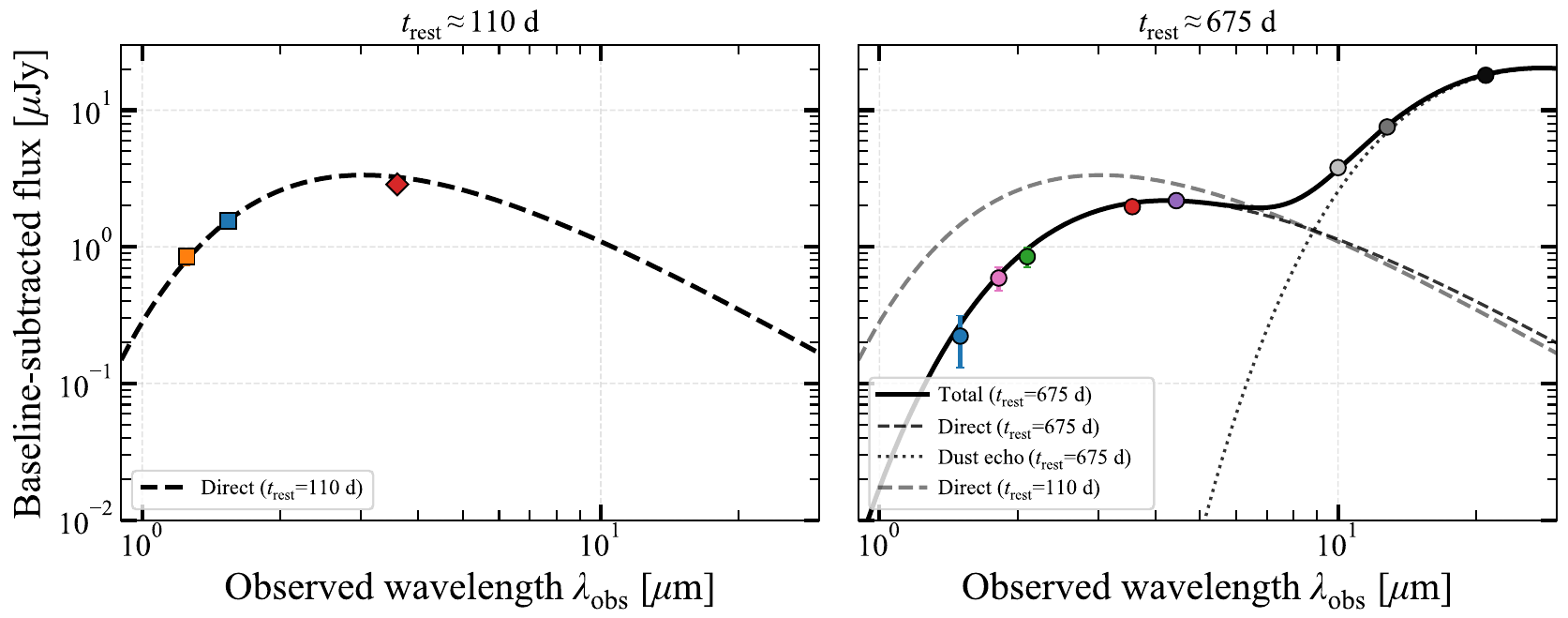}
\end{center}
\vspace{-0.2cm}
\caption{\textbf{SEDs at near peak and at late epoch.} The color symbols and the zero of the rest-frame date are the same as Figure~\ref{fig:light_curve}. 
The zero-point of the rest-frame time is defined identically to Figure~\ref{fig:light_curve}. 
\textbf{Left:} Observed-frame spectral energy distribution (SED) at $t_{\rm rest}\approx110$~d, constructed from baseline-subtracted photometry in \textit{HST}/F125W and F160W and \textit{Spitzer}/IRAC 3.6$\mu$m. The dashed black curve shows the best-fit \emph{direct} transient component from the best-fit light-curve model (Figure~\ref{fig:light_curve}). 
\textbf{Right:} Corresponding SED at $t_{\rm rest}\approx675$~d from baseline-subtracted \textit{JWST}/NIRCam (F150W, F182M, F210M, F356W, F444W) and \textit{JWST}/MIRI (F1000W, F1280W, F2100W) photometry. The solid black curve shows the best-fit \emph{total} model (direct transient + dust echo; Eq.~\ref{eq:f_tot}), decomposed into the direct component (grey dashed) and the dust-echo component (black dotted; Figure~\ref{fig:dust_echo}). For reference, the $t_{\rm rest}\approx110$~d direct-component model is overplotted in the right panel (grey dash-dotted) to highlight the cooling of the UV--optical continuum with time. 
}
\label{fig:seds}
\end{figure*}

To visualize the evolution of the transient reprocessed energy and the emergence of dust-reprocessed emission, we constructed two representative SEDs, which are plotted in Figure~\ref{fig:seds}. We selected epochs near the fitted peak but with multiple observed data points ($t_{\rm rest}\approx110$~d) and at a late epoch ($t_{\rm rest}\approx675$~d). For each epoch, we display baseline-subtracted photometric measurements drawn from the best-fit multi-band light-curve model (Section~\ref{sec:lc_fit}), choosing data points closest in rest-frame time to the target epoch in the relevant bands.
Model SED curves are computed from the best-fit parameters derived from the multi-band light-curve model for the direct component, and from the dust-echo model (Section~\ref{sec:dust_echo}) for the mid-infrared component. 

As shown in Figure~\ref{fig:seds}, the same best-fit light-curve and dust-echo models presented in Figure~\ref{fig:light_curve} provide an excellent and self-consistent description of the multi-band SEDs at both epochs without introducing any additional free parameters. 
The direct component naturally accounts for the hot UV--optical continuum near peak and its subsequent cooling at later times, while the delayed dust-reprocessed component reproduces the emergence of the mid-infrared excess at late epochs.
Importantly, these simple physically motivated models (TDE + dust echo) simultaneously explain the observed flux evolution across multiple bands and multiple epochs, in contrast to the alternative scenarios of SNs or stochastic AGN variability (Appendix~\ref{sec:alternative}). 

\subsection{Estimate of the disrupted-star mass}
The agreement with the TDE light-curve model allows us to estimate the physical properties of the event. The best-fit model implies a characteristic temperature of $T_{\rm peak}\simeq1.4\times10^{4}$~K, a peak bolometric luminosity of $L_{\rm bol}\simeq3\times10^{45}$~erg~s$^{-1}$, and a total radiated energy of $\simeq1.5\times10^{53}$~erg. Integrating the bolometric light curve and adopting standard fallback energetics, these values imply a disrupted-star mass of order a few solar masses at least (see below).

The inferred temperature is typical of local TDEs \citep{gezari2021}, while the luminosity and radiated energy place this event among the most energetic TDEs known \citep{hinkle2025}. Statistical tests indicate that such smooth, multi-band variability is extremely unlikely ($\lesssim10^{-5}$) under stochastic AGN variability, while SLSNe are disfavored by the observed duration and energetics (see Appendix~\ref{sec:alternative}). 

\label{sec:star_mass}

We estimate the mass of the disrupted star by equating the total radiated energy of the transient component to the rest-mass energy released by accretion during a TDE. 
The baseline-subtracted bolometric light curve is described by a smooth broken power-law decay with a late-time slope fixed to $t^{-5/3}$, as expected for fallback accretion. 
Integrating the best-fit bolometric luminosity over time therefore yields the total radiated energy,
\begin{equation}
E_{\rm rad} \equiv \int L_{\rm bol}(t)\,dt \simeq (1\text{--}2)\times10^{53}\ {\rm erg},
\end{equation}
where the quoted range reflects the integration limits constrained by the data and the uncertainty in extrapolating the late-time decay.

Assuming a radiative efficiency $\eta$, the corresponding accreted mass is
\begin{equation}
M_{\rm acc} = \frac{E_{\rm rad}}{\eta c^{2}}.
\end{equation}
In a TDE, approximately half of the stellar mass becomes unbound during the disruption, implying a lower limit on the progenitor mass of
\begin{equation}
m_\star \gtrsim 2\,M_{\rm acc}.
\end{equation}
Adopting a canonical efficiency $\eta=0.1$ yields $m_\star\gtrsim1$--$2\,M_\odot$, while allowing for a lower effective efficiency ($\eta\simeq0.03$), appropriate for super-Eddington accretion with strong photon trapping and outflows, increases the inferred mass to $m_\star\gtrsim3$--$5\,M_\odot$. 
Recent broadband SED modeling of TDEs suggests that the radiative efficiency may be time-dependent, with lower values at early, super-Eddington phases and higher values at later times \citep{rli2024}, 
consistent with the range of efficiencies considered here.
We therefore conservatively conclude that the disrupted star had a mass of order a few solar masses, noting that additional uncertainties arise from the efficiency of circularization and any residual AGN contribution to the transient luminosity.

As an independent sanity check, we also estimate the stellar mass required for a tidal disruption to occur given the black-hole mass of \targ. 
The tidal radius is defined as
\begin{equation}
 r_{\rm tidal}=\left(\eta_\star^2 \frac{M_{\rm BH}}{m_\star}\right)^{1/3}r_\star,
\end{equation}
where $M_{\rm BH}$ is the black-hole mass, $m_\star$ and $r_\star$ are the stellar mass and radius, and $\eta_\star\simeq0.84$ depends on the stellar internal structure. 
For a tidal disruption to occur outside the event horizon, the tidal radius must exceed the radius of the innermost stable circular orbit (ISCO), $r_{\rm tidal}/R_{\rm ISCO}>1$. 
For a non-spinning black hole this condition translates approximately to
\begin{equation}
M_{\rm BH,8}^{-2/3} m_{\star,0}^{-1/3} r_{\star,0} \gtrsim 3,
\end{equation}
where $M_{\rm BH,8}=M_{\rm BH}/10^{8}M_\odot$, $m_{\star,0}=m_\star/M_\odot$, and $r_{\star,0}=r_\star/R_\odot$. 
For main-sequence stars the mass--radius relation can be approximated as $r_\star\propto m_\star^{\xi}$ with $\xi\simeq0.6$--$0.8$. 
Adopting $\xi\simeq2/3$, appropriate for intermediate-mass stars, yields a minimum stellar mass of
\begin{equation}
m_\star \gtrsim 3\,M_\odot
\end{equation}
for $M_{\rm BH}\simeq3.5\times10^{7}M_\odot$. 
This estimate is broadly consistent with the stellar mass inferred from the energetics above. 
Such stellar masses correspond to B-type stars with lifetimes of $\lesssim200$--$300$\,Myr, which are naturally expected in the young, intensely star-forming host galaxy of \targ. Specifically, using the current star-formation rate and stellar mass (${\rm SFR}\sim300\,M_\odot\,{\rm yr^{-1}}$, $M_\star\sim3\times10^{10}\,M_\odot$) to estimate a characteristic stellar assembly timescale yields a value of order $\sim100$\,Myr, broadly consistent with the presence of such intermediate-mass stars.
We therefore conclude that the inferred stellar mass required for a tidal disruption is consistent with both the energetics of the transient and the stellar population of the host galaxy.

\section{Discussion}
\label{sec:discussion}

\subsection{Alternative scenarios}
\label{sec:disc_alt}

In Appendix~\ref{sec:alternative}, we test two alternatives against the observed light curve, as summarized in Figure~\ref{fig:lc_comp}.
A scaled version of the slowly declining SLSN SN2015bn fades well before the rest-frame $\sim 2$~yr decline measured here, as expected from the $\lesssim 1$~yr decline times of the class \citep{galyam2019}, and falls an order of magnitude short of the radiated energy of $\sim 1.5 \times 10^{53}$~erg.
Stochastic disk variability fares no better.
Among $2 \times 10^{5}$ damped random walk realizations drawn from the priors of \citet{macleod2010}, the smooth, coherent multi-band evolution observed here is reproduced with probability $\lesssim 10^{-5}$.
A TDE therefore remains the most natural interpretation of the light curve, energetics, and dust echo taken together, although confirmation over a longer spectroscopic or photometric baseline is still needed, and continued monitoring will provide it.

Even if the transient proves not to be a TDE, either surviving alternative would be remarkable.
A supernova fading over a rest-frame $\sim 2$~yr and radiating $\sim 1.5 \times 10^{53}$~erg would lie beyond the known SLSN population \citep{galyam2019}, an exceptional stellar explosion whose energy scale is otherwise approached mainly by the ENTs \citep{hinkle2025}.
Accretion-disk variability would be no less striking.
GNz7q accretes at $\lambda_{\rm Edd} = 2.7$ \citep{fei2026}, so this option would make it a super-Eddington quasar whose rest-frame ultraviolet luminosity changed by $\sim 1.1$~mag, a factor of $\sim 3$, within $\sim 2$~yr, a drastic accretion change at the epoch of reionization.
The same monitoring will distinguish between these outcomes.

\subsection{More tidal disruption events in the early Universe?}
\label{sec:disc_rates}

\begin{figure}
\begin{center}
\includegraphics[angle=0,width=\columnwidth]{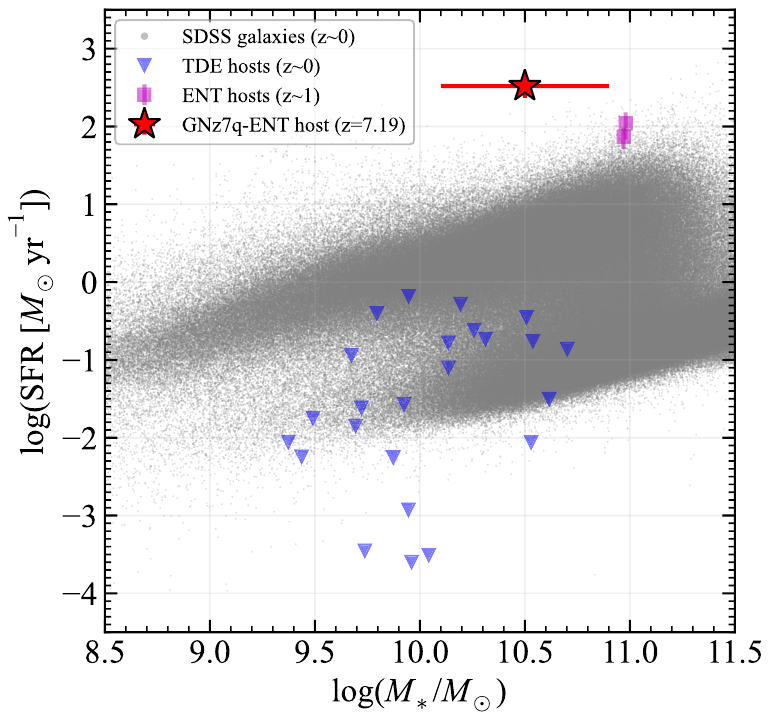}
\end{center}
\vspace{-0.2cm}
\caption{\small
\textbf{Host-galaxy context of GNz7q-ENT.}
Host-galaxy properties of GNz7q-ENT compared to those of TDE hosts in the local Universe and extreme nuclear transients (ENTs) at $z\sim1$.
In contrast to local TDEs, the hosts of the most energetic nuclear transients are massive, dusty, and intensely star-forming systems with high $M_{\rm star}$ and SFR, which is in line with the high stellar-density environments may naturally enhance the incidence of ENTs, extending this emerging connection to the epoch of reionization.
}
\label{fig:host}
\end{figure}

In the local Universe, TDEs are preferentially found in post-starburst galaxies rather than in actively star-forming systems \citep{french2020}. 
Nevertheless, the extreme transient observed in \targ\ is physically self-consistent given the nature of its host galaxy. 
\targ\ resides in a massive, dusty galaxy with a very compact dust continuum ($r_{\rm e}<480$~pc; \citealt{fujimoto2022}), indicative of a dense nuclear star-forming environment, and hosts a relatively low-mass black hole ($M_{\rm BH}<10^{8}\,M_\odot$; \citealt{fei2026}), a regime in which stellar tidal disruptions are theoretically and observationally expected to be most efficient \citep{stone2016a,french2020}. 
Recent studies have also shown that the most energetic TDEs at $z\sim1$ are associated with massive, dusty star-forming galaxies \citep{hinkle2025}. Figure~\ref{fig:host} places the host of \ent\ alongside local TDE hosts and those of the ENTs at $z\sim1$. The event reported here therefore extends this connection between ENTs and dense, star-forming environments to the epoch of reionization.

\ent\ belongs to the most energetic class of nuclear flares known, occupying the extreme tail of the luminosity--duration distribution compared to canonical TDEs and SLSNe \citep{hinkle2025}. Despite its exceptional energetics, the event exhibits a coherent set of observational properties, including its light-curve evolution, duration, and additional broad-line component, that are naturally explained by a TDE. 
TDEs are intrinsically rare in the local Universe \citep{stone2016b,van-velzen2021}. The serendipitous detection of \ent\ in a $z=7.19$ quasar, enabled by the long temporal baseline in the GOODS-North field, suggests that such events may be more readily observable in the early Universe. The physical conditions in high-redshift galaxies, including high stellar densities \citep[e.g.,][]{adamo2024,vanzella2023,mowla2024,fujimoto2024} and abundant low-mass black holes \citep{fei2025}, may enhance both the incidence and observability of transient accretion episodes. \ent\ may therefore represent an example of short-lived, intense fueling events during the earliest phases of black-hole and galaxy assembly.

\subsection{From the nuclear event to circumnuclear dust and cosmology}
\label{sec:disc_cosmo}

Beyond its classification, this discovery points to TDEs as a probe of black-hole growth in the early Universe. Because TDEs can occur around otherwise faint or undetectable black holes, they provide direct access to episodic accretion and may offer a complementary view of the low-mass black-hole population at high redshift. 
In addition, the detection of a mid-infrared dust echo demonstrates that time-domain observations can trace the response of circumnuclear dust to rapid accretion events even at $z>7$. Such measurements make it possible to probe the structure and evolution of the nuclear environment around early black holes. 
Lastly, AGN dust reverberation has been proposed as a distance indicator, suggesting that dust echoes from high-redshift accretion events could ultimately provide a complementary cosmological probe \citep{yoshii2014,honig2014}, within the first billion years of cosmic history.
This result demonstrates that such events reveal a previously unexplored regime of time-domain black-hole growth and its dusty nuclear environment in the early Universe, with potential implications for future cosmological applications. Wide-field time-domain surveys with \textit{Roman} and \textit{Euclid} in the near-infrared, complemented by LSST at lower redshifts, are expected to enlarge the census of such nuclear transients at high redshift \citep{karmen2026}, making them a systematic probe of black-hole growth in the reionization era. 

\section*{Acknowledgments}

We thank Kotaro Kohno, Adam Muzzin, Jinyi Yang, Feige Wang, Fengwu Sun, and Wenkai Li for valuable feedback on the interpretations of this transient event in \targ. 
This work is based on observations made with the NASA/ESA/CSA James Webb Space Telescope (program ID: 1895, 4762, 5407, 7404, 9214), and Subaru (program ID: S07A-010), \textit{Spitzer} (program ID: 169, 61040, 80215, 11134), and \textit{HST} (program ID: 12443, 12444, 12445). 
For \textit{JWST}, the data were obtained from the Mikulski Archive for Space Telescopes at the Space Telescope Science Institute, which is operated by the Association of Universities for Research in Astronomy, Inc., under NASA contract NAS 5-03127 for \textit{JWST}. 

\paragraph*{Funding:}
Support for program \#4762 was provided by NASA through a grant from the Space Telescope Science Institute, which is operated by the Association of Universities for Research in Astronomy, Inc., under NASA contract NAS 5-03127.
S.F. acknowledges support from the Dunlap Institute, funded through an endowment established by the David Dunlap family and the University of Toronto.
K.I. and L.C.H. acknowledge support from the National Natural Science Foundation of China (12573015, W2532003, 12233001), the Beijing Natural Science Foundation (IS25003), and the China Manned Space Program (CMS-CSST-2025-A09).
M.O. is supported by the Japan Society for the Promotion of Science (JSPS) KAKENHI grant No. JP24K22894. 
PGP-G acknowledges support from grant PID2022-139567NB-I00 funded by Spanish Ministerio de Ciencia, Innovaci\'on y Universidades MCIU/AEI/10.13039/501100011033,
FEDER {\it Una manera de hacer
Europa}. 
L.C. acknowledges support from grant PID2021-121788NB-I00 funded by Spanish Ministerio de Ciencia, Innovaci\'on y Universidades MCIU/AEI/10.13039/501100011033, FEDER {\it Una manera de hacer
Europa.}. 
P.O., R.A.M., M.X. acknowledge support from the Swiss State Secretariat for Education, Research and Innovation (SERI) under contract number MB22.00072, as well as the Swiss National Science Foundation (SNSF) through project grant 200020\_207349. 
J.A.-M. acknowledges support by grants PID2024-158856NA-I00 \& PIB2021-127718NB-I00 from the Spanish Ministry of Science and Innovation/State Agency of Research MCIN/AEI/10.13039/501100011033 and by ``ERDF A way of making Europe''.

\paragraph*{Data and materials availability:}
This paper makes use of the \jwst\ data mainly from \#GO-4762, together with \#GO-1895, \#GO-5407, and \#GO-7404 that are all available at  \url{https://archive.stsci.edu/}.  
The reduced \jwst\ NIRCam images are available at \url{https://dawn-cph.github.io/dja/}.
Other datasets generated and/or analyzed during the current study are available from the corresponding author upon reasonable request.
The NIRCam data were processed with {\sc grizli} available at \url{https://github.com/gbrammer/grizli}. 
The detector variation map is available at \url{https://github.com/gbrammer/grizli/pull/107}. 

\paragraph*{Use of generative AI:}
We acknowledge the use of a generative AI tool to refine the text and the code used in this work. 
All figures and quoted values were reproduced and verified by the authors.

\appendix
\renewcommand{\theHfigure}{\thesection\arabic{figure}}
\renewcommand{\theHequation}{\thesection\arabic{equation}}

\section{Testing Photometric Systematics Using Nearby Sources}
\label{sec:test}

To assess whether the observed variability of \targ\ could arise from photometric systematics rather than intrinsic astrophysical variability, we performed a set of control tests using nearby, isolated sources within the same fields. 
We selected four comparison objects located close to \targ\ that are moderately compact and free from blending across all relevant wavelengths.

For each comparison source, we measured multi-epoch photometry following exactly the same procedures adopted for \targ, including identical aperture sizes, aperture corrections, signal-to-noise cuts, and the same combinations of instruments and filters (\textit{HST}, \textit{Spitzer}/IRAC, and \textit{JWST}/NIRCam). 
We specifically examined flux variations in matched filter pairs probing similar rest-frame wavelengths (e.g., F160W--F150W, F125W--F115W, IRAC 3.6~$\mu$m--F356W, and IRAC 4.5~$\mu$m--F444W), normalizing each light curve by its mean flux to facilitate direct comparison across instruments.

\begin{figure*}[p]
\begin{center}
\includegraphics[angle=0,width=1.0\textwidth]{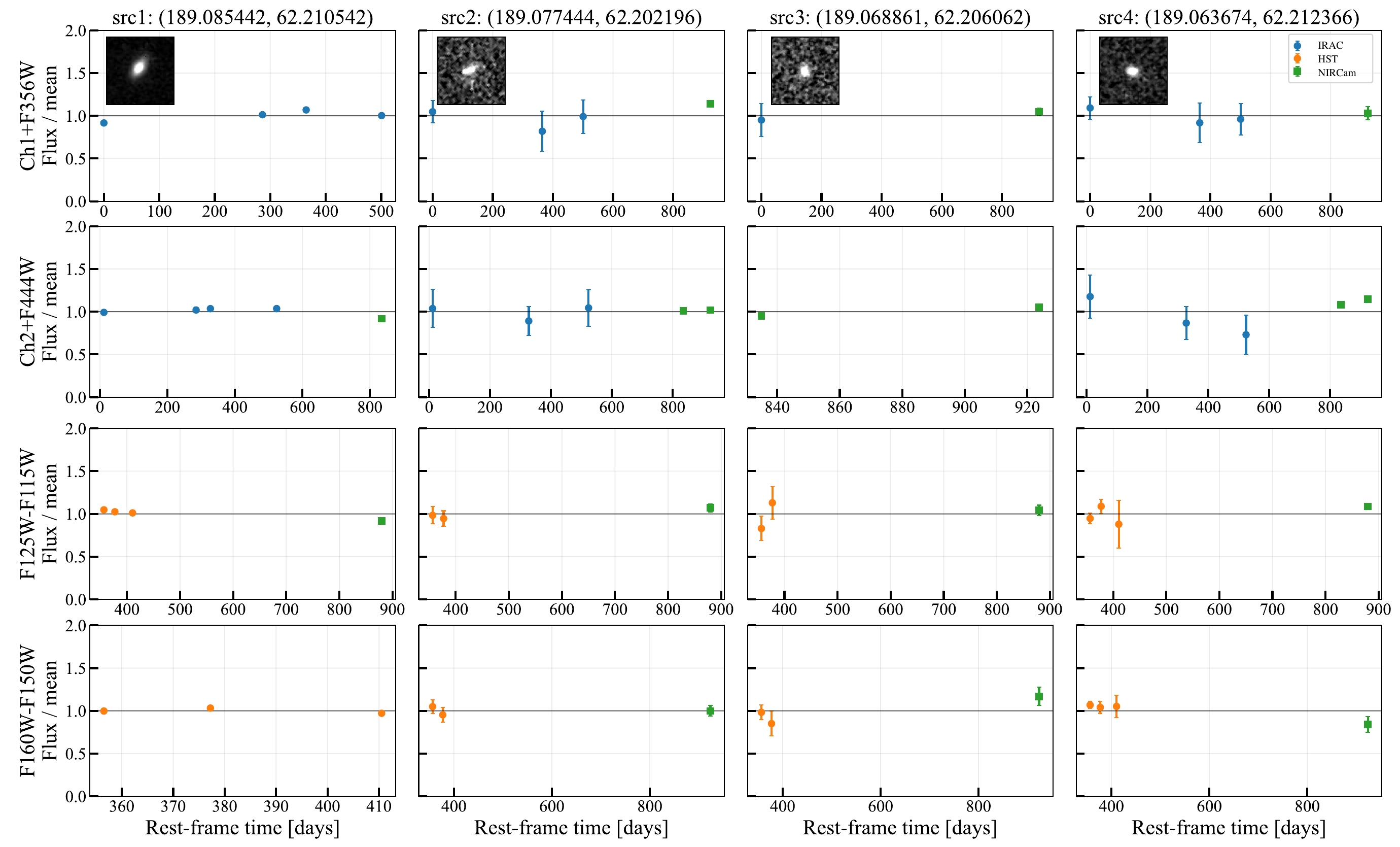}
\end{center}
\vspace{-0.2cm}
\caption{\small
\textbf{Flux-variation tests using nearby comparison sources.}
Normalized multi-epoch photometry for four isolated comparison objects located near \targ, measured using the same photometric procedures adopted for GNz7q. 
For each source, fluxes are shown in matched filter pairs probing similar rest-frame wavelengths: IRAC 3.6~$\mu$m with NIRCam/F356W, IRAC 4.5~$\mu$m with NIRCam/F444W, \textit{HST}/F125W with NIRCam/F115W, and \textit{HST}/F160W with NIRCam/F150W. 
All light curves are normalized by their mean flux, and only measurements with ${\rm S/N}>3$ are shown. 
Colors and symbols indicate different instruments (\textit{Spitzer}/IRAC, \textit{HST}, and \textit{JWST}/NIRCam). 
None of the comparison sources shows coherent long-term or wavelength-dependent variability, demonstrating that the large-amplitude, chromatic variability observed in GNz7q is not driven by photometric systematics.
}
\label{fig:test}
\end{figure*}

Figure~\ref{fig:test} shows the resulting normalized light curves for these comparison sources, including only measurements with ${\rm S/N}>3$. 
None of the control objects exhibits statistically significant long-term or wavelength-dependent variability comparable to that observed in \targ. 
The absence of correlated variability in these sources demonstrates that the observed behavior of GNz7q cannot be attributed to instrumental differences, filter-response mismatches, PSF variations, or details of the photometric extraction, and instead reflects genuine intrinsic variability of the source.

\section{Alternative Scenarios}
\label{sec:alternative}

While we confirm the excellent agreement between the smooth, multi-band light curve observed in \ent\ and the TDE + dust echo modeling (see Figure~\ref{fig:light_curve} and Figure~\ref{fig:seds}), we also quantify how well alternative explanations could reproduce the observed multi-band behavior of \ent. We carried out two independent, observationally anchored tests: (1) a direct comparison to a representative slow-evolving superluminous supernova (SLSN) light curve, and (2) a Monte Carlo test of stochastic AGN-like variability modeled as a damped random walk (DRW; Ornstein--Uhlenbeck process) using the empirical MacLeod et\,al.\ (2010) priors.

First for the SLSN comparison, we selected one of the best-studied SLSNe, SN2015bn \citep{nicholl2016}, as a representative long-lived SLSN template, adopting light-curve data from the Open Supernova Catalog \citep{guilochon2017}. For a direct shape comparison, we (i) selected bands whose rest-frame wavelengths match \ent\ rest-frame UV ($\approx 1800$\,\AA) and optical ($\approx 5500$\,\AA) windows, (ii) converted the SN photometry into comparable flux units and rest-frame times, and (iii) normalized and shifted the SN template so that its peak flux and rest-frame peak time match those of \ent\ separately in the UV-like and optical-like bands. No additional temporal stretching or free parameters were introduced.

\begin{figure*}[t!]
\begin{center}
\includegraphics[angle=0,width=1.0\textwidth]{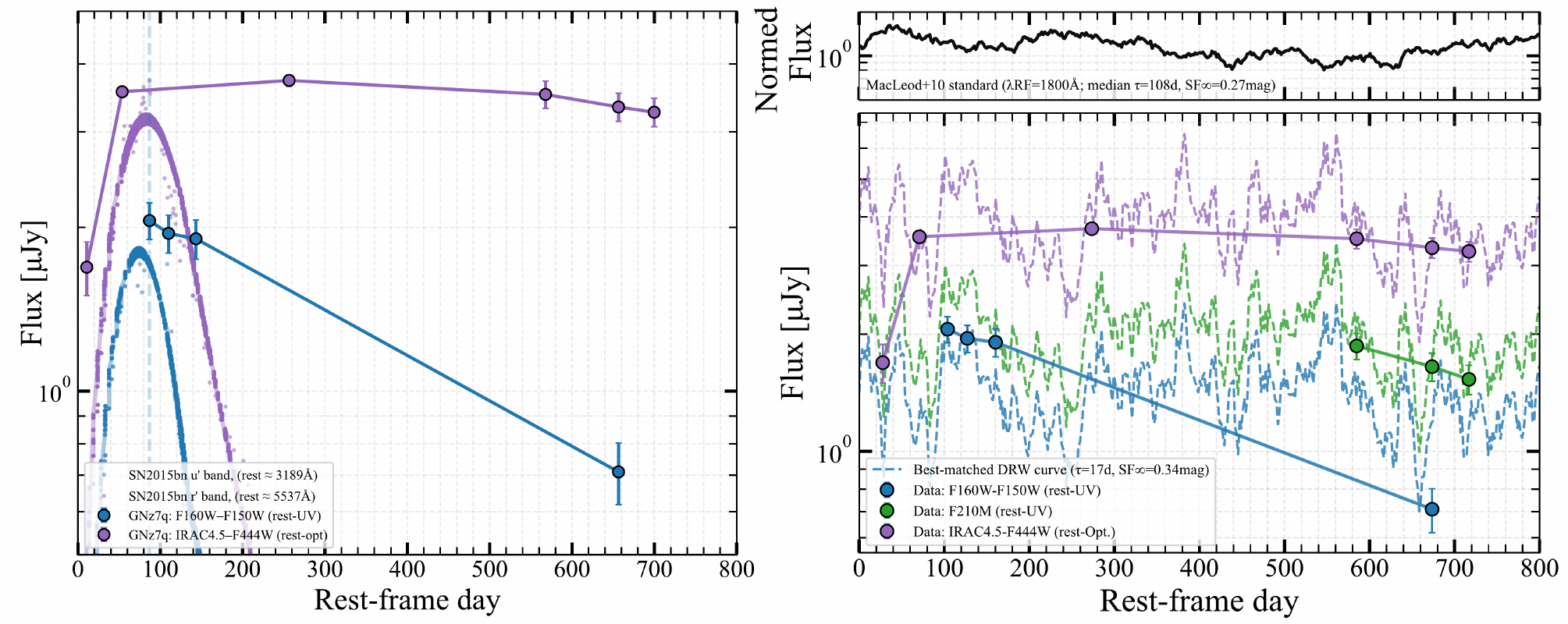}
\end{center}
\vspace{-0.2cm}
\caption{\textbf{Comparison of \ent\ light curve with representative SLSN and DRW/AGN variability.}
\textbf{Left:} \ent's rest-frame UV (F160W--F150W; blue) and optical (IRAC4.5--F444W; purple) light curves, in the same color coding as Figure~\ref{fig:light_curve}.  
A representative slow-evolving SLSN (SN2015bn; \citealt{nicholl2016}) is shown for comparison after normalizing the SN peak fluxes to match GNz7q in the rest-UV and optical bands separately (see Appendix~\ref{sec:alternative}). 
\textbf{Right:} The top panel shows a single example `standard' Damped Random Walk (DRW) realization drawn from the median \citet{macleod2010} priors (median $\tau$, SF$_\infty$ indicated in the inset). The bottom panel compares the \ent\ data (same color coding as Figure~\ref{fig:light_curve}) to the DRW-based variability tests: the dashed curves represent the best-matched DRW realization selected from a 200,000 Monte Carlo ensemble with
the MacLeod priors ($\tau$ and SF$_\infty$ scatters) and constrained to pass a \ent-like  event light curve (see Appendix~\ref{sec:alternative}). The probability of obtaining the \ent-like smooth multi-band light curves, due to the limited temporal baseline under the AGN/DRW-like variability, is vanishingly small ($\ll 10^{-3}\%$). 
}
\label{fig:lc_comp}
\end{figure*}

The left panel of Figure~\ref{fig:lc_comp} shows the resulting comparison. Even after amplitude scaling, the SLSN template declines substantially faster than \ent\ and fails to reproduce the observed duration (rest-frame $\sim$2\,yr). Furthermore, the implied integrated radiated energy and peak luminosity remain inconsistent with typical SLSNe by an order of magnitude. The combined discrepancies in duration and energetics make a canonical SLSN origin unlikely \citep{hinkle2025}.

Second for the DRW/AGN variability, We next evaluated whether the smooth, multi-band evolution of \ent\ could arise from stochastic AGN variability given the limited temporal baseline. We adopted the \citet{macleod2010} scaling relations to construct log-normal priors for the characteristic damping time $\tau$ and asymptotic structure function SF$_\infty$, as functions of rest-frame wavelength, host absolute magnitude, and black-hole mass. We adopted intrinsic scatters of 0.30~dex in $\log_{10}(\tau)$ and 0.20~dex in $\log_{10}({\rm SF}_\infty)$, consistent with MacLeod et\,al.\ (2010).

We performed the following Monte Carlo experiment. A probability test draws 200,000 realizations with independent DRW parameter sets sampled from the MacLeod priors, simulates a joint latent DRW light curve evaluated at the observed sampling times of \ent, and measures the probability $P_{\rm smooth}$ that the noisy, sampled realization satisfies a predefined \ent-like event definition. We define an \ent-like event by the following criteria: 
(a) a rest-frame UV decline of at least 1.0\,mag between 550 and 650 rest-frame days after the UV peak; 
(b) simultaneously small rest-frame optical variation (peak-to-peak $\leq 0.2$\,mag) within the same interval; and 
(c) monotonicity constraints allowing at most 0 brightenings in the UV and at most 1 brightening in the optical band after peak. These criteria capture the core temporal features of \ent\ while remaining conservative in allowing measurement noise to assist passing realizations.

For simplicity and conservativeness, this test uses two representative bands per wavelength regime: \textit{HST}/F160W and NIRCam/F150W for the rest-UV, and \textit{Spitzer}/IRAC 3.6$\mu$m and NIRCam/F356W for the rest-optical. Under the adopted priors and sampling pattern, we obtain $P_{\rm smooth}\simeq 10^{-5}$. We note that this estimate is still conservative: the observed \ent\ light curve exhibits coherent evolution across more filters and epochs than imposed in the above definition. Requiring the same monotonic multi-band behavior simultaneously across all available filters would further suppress the probability relative to $P_{\rm smooth}$ (i.e., $\ll10^{-5}$).

The right panel of Figure~\ref{fig:lc_comp} presents a representative ``best-matched'' AGN/DRW realization drawn from the Monte Carlo ensemble. We first select realizations satisfying criteria (a)--(c), then rank them by $\chi^2$ relative to the observed multi-band data. For visualization, we compute $\chi^2$ using only \textit{HST}/F160W, NIRCam/F150W, F210M, F356W, and \textit{Spitzer}/IRAC 3.6$\mu$m, and display the corresponding curves at 1.5$\mu$m, 2.1$\mu$m, and 3.6$\mu$m. While parts of the \ent\ light curve can be approximated under highly tuned DRW parameters, reproducing the full multi-band coherence requires extreme fine-tuning and remains statistically disfavored, consistent with the low probability derived above.

Both the SLSN template and stochastic AGN variability scenarios struggle to reproduce the coherent, multi-band light-curve evolution of \ent. These tests therefore support the interpretation that a TDE-like, self-consistent light-curve plus dust-echo model provides a more natural explanation of the multi-band, multi-epoch observations of \ent.

\bibliographystyle{aasjournal}
\bibliography{reference}

\begin{thebibliography}{}
\expandafter\ifx\csname natexlab\endcsname\relax\def\natexlab#1{#1}\fi
\providecommand{\url}[1]{\href{#1}{#1}}
\providecommand{\dodoi}[1]{doi:~\href{http://doi.org/#1}{\nolinkurl{#1}}}
\providecommand{\doeprint}[1]{\href{http://ascl.net/#1}{\nolinkurl{http://ascl.net/#1}}}
\providecommand{\doarXiv}[1]{\href{https://arxiv.org/abs/#1}{\nolinkurl{https://arxiv.org/abs/#1}}}

\bibitem[{{Adamo} {et~al.}(2024){Adamo}, {Bradley}, {Vanzella}, {Claeyssens},
  {Welch}, {Diego}, {Mahler}, {Oguri}, {Sharon}, {Abdurro'uf}, {Hsiao}, {Xu},
  {Messa}, {Lassen}, {Zackrisson}, {Brammer}, {Coe}, {Kokorev}, {Ricotti},
  {Zitrin}, {Fujimoto}, {Inoue}, {Resseguier}, {Rigby}, {Jim{\'e}nez-Teja},
  {Windhorst}, {Hashimoto}, \& {Tamura}}]{adamo2024}
{Adamo}, A., {Bradley}, L.~D., {Vanzella}, E., {et~al.} 2024, \nat, 632, 513,
  \dodoi{10.1038/s41586-024-07703-7}

\bibitem[{{Anderson} {et~al.}(2014){Anderson}, {Gonz{\'a}lez-Gait{\'a}n},
  {Hamuy}, {Guti{\'e}rrez}, {Stritzinger}, {Olivares E.}, {Phillips},
  {Schulze}, {Antezana}, {Bolt}, {Campillay}, {Castell{\'o}n}, {Contreras}, {de
  Jaeger}, {Folatelli}, {F{\"o}rster}, {Freedman}, {Gonz{\'a}lez}, {Hsiao},
  {Krzemi{\'n}ski}, {Krisciunas}, {Maza}, {McCarthy}, {Morrell}, {Persson},
  {Roth}, {Salgado}, {Suntzeff}, \& {Thomas-Osip}}]{anderson2014}
{Anderson}, J.~P., {Gonz{\'a}lez-Gait{\'a}n}, S., {Hamuy}, M., {et~al.} 2014,
  \apj, 786, 67, \dodoi{10.1088/0004-637X/786/1/67}

\bibitem[{{Andreoni} {et~al.}(2022)}]{andreoni2022}
{Andreoni}, I., {et~al.} 2022, \nat, 612, 430,
  \dodoi{10.1038/s41586-022-05465-8}

\bibitem[{{Ashby} {et~al.}(2013){Ashby}, {Stanford}, {Brodwin}, {Gonzalez},
  {Martinez-Manso}, {Bartlett}, {Benson}, {Bleem}, {Crawford}, {Dey},
  {Dressler}, {Eisenhardt}, {Galametz}, {Jannuzi}, {Marrone}, {Mei}, {Muzzin},
  {Pacaud}, {Pierre}, {Stern}, \& {Vieira}}]{ashby2013}
{Ashby}, M.~L.~N., {Stanford}, S.~A., {Brodwin}, M., {et~al.} 2013, \apjs, 209,
  22

\bibitem[{{Ashby} {et~al.}(2015){Ashby}, {Willner}, {Fazio}, {Dunlop}, {Egami},
  {Faber}, {Ferguson}, {Grogin}, {Hora}, {Huang}, {Koekemoer}, {Labb{\'e}}, \&
  {Wang}}]{ashby2015}
{Ashby}, M.~L.~N., {Willner}, S.~P., {Fazio}, G.~G., {et~al.} 2015, \apjs, 218,
  33, \dodoi{10.1088/0067-0049/218/2/33}

\bibitem[{{Ba{\~n}ados} {et~al.}(2018){Ba{\~n}ados}, {Venemans},
  {Mazzucchelli}, {Farina}, {Walter}, {Wang}, {Decarli}, {Stern}, {Fan},
  {Davies}, {Hennawi}, {Simcoe}, {Turner}, {Rix}, {Yang}, {Kelson}, {Rudie}, \&
  {Winters}}]{banados2018}
{Ba{\~n}ados}, E., {Venemans}, B.~P., {Mazzucchelli}, C., {et~al.} 2018, \nat,
  553, 473, \dodoi{10.1038/nature25180}

\bibitem[{{Barvainis}(1987)}]{barvainis1987}
{Barvainis}, R. 1987, \apj, 320, 537, \dodoi{10.1086/165571}

\bibitem[{{Baskin} \& {Laor}(2018)}]{baskin2018}
{Baskin}, A., \& {Laor}, A. 2018, \mnras, 474, 1970,
  \dodoi{10.1093/mnras/stx2850}

\bibitem[{{Bosman} {et~al.}(2025){Bosman}, {{\'A}lvarez-M{\'a}rquez}, {Davies},
  {Protu{\v{s}}ov{\'a}}, {Hennawi}, {Yang}, {Spina}, {Colina}, {Fan},
  {{\"O}stlin}, {Walter}, {Wang}, {Ward}, {Alonso Herrero}, {Barth},
  {Belladitta}, {Boogaard}, {Caputi}, {Connor},
  {{\v{D}}urov{\v{c}}{\'\i}kov{\'a}}, {Eilers}, {Crespo G{\'o}mez}, {Hjorth},
  {Jun}, {Langeroodi}, {Liu}, {Lupi}, {Mazzucchelli}, {Pye}, {Rinaldi}, {van
  der Werf}, \& {Volonteri}}]{bosman2025}
{Bosman}, S. E.~I., {{\'A}lvarez-M{\'a}rquez}, J., {Davies}, F.~B., {et~al.}
  2025, arXiv e-prints, arXiv:2511.02902, \dodoi{10.48550/arXiv.2511.02902}

\bibitem[{{Calzetti} {et~al.}(2000){Calzetti}, {Armus}, {Bohlin}, {Kinney},
  {Koornneef}, \& {Storchi-Bergmann}}]{calzetti2000}
{Calzetti}, D., {Armus}, L., {Bohlin}, R.~C., {et~al.} 2000, \apj, 533, 682,
  \dodoi{10.1086/308692}

\bibitem[{{Coulter} {et~al.}(2026){Coulter}, {Larison}, {Pierel}, {Fujimoto},
  {Kokorev}, {Allingham}, {Moriya}, {Siebert}, {Asada}, {Bezanson},
  {Brada{\v{c}}}, {Brammer}, {Chisholm}, {Coe}, {Dayal}, {Engesser},
  {Finkelstein}, {Fox}, {Furtak}, {Koekemoer}, {Moore}, {Nakane}, {Ouchi},
  {Pan}, {Quimby}, {Rest}, {Richard}, {Robbins}, {Strolger}, {Sun}, {Treu},
  {Yanagisawa}, {Abdurro'uf}, {Agrawal}, {Amor{\'\i}n}, {Anderson}, {Angulo},
  {Atek}, {Bauer}, {Bradley}, {Bromm}, {Bronikowski}, {Conselice}, {DeCoursey},
  {DerKacy}, {Desprez}, {Dhawan}, {Diego}, {Egami}, {Faisst}, {Frye}, {Gomez},
  {Gonz{\'a}lez-Otero}, {Griggio}, {Harikane}, {Inayoshi}, {Jha},
  {Jim{\'e}nez-Teja}, {Kartaltepe}, {Kelly}, {Kwok}, {Lane}, {Li}, {Lobbe},
  {Lopes}, {Lucas}, {Magdis}, {Martis}, {Matthee}, {Meena}, {Naidu}, {Noirot},
  {Oguri}, {Padilla Gonzalez}, {Pascale}, {Petrushevska}, {Ricotti},
  {Schaerer}, {Schuldt}, {Shahbandeh}, {Sheu}, {Shukawa}, {Tsujita},
  {Vanzella}, {Wang}, {Weaver}, {Williams}, {Windhorst}, {Xu}, {Zenati}, \&
  {Zitrin}}]{coulter2026}
{Coulter}, D.~A., {Larison}, C., {Pierel}, J. D.~R., {et~al.} 2026, arXiv
  e-prints, arXiv:2601.04156, \dodoi{10.48550/arXiv.2601.04156}

\bibitem[{{De Cia} {et~al.}(2018){De Cia}, {Gal-Yam}, {Rubin}, {Leloudas},
  {Vreeswijk}, {Perley}, {Quimby}, {Yan}, {Sullivan}, {Fl{\"o}rs}, {Sollerman},
  {Bersier}, {Cenko}, {Gal-Yam}, {Maguire}, {Ofek}, {Prentice}, {Schulze},
  {Spyromilio}, {Valenti}, {Arcavi}, {Corsi}, {Howell}, {Mazzali}, {Kasliwal},
  {Taddia}, \& {Yaron}}]{decia2018}
{De Cia}, A., {Gal-Yam}, A., {Rubin}, A., {et~al.} 2018, \apj, 860, 100,
  \dodoi{10.3847/1538-4357/aab9b6}

\bibitem[{{DeCoursey} {et~al.}(2025){DeCoursey}, {Egami}, {Pierel}, {Sun},
  {Rest}, {Coulter}, {Engesser}, {Siebert}, {Hainline}, {Johnson},
  {et~al.}}]{decoursey2025}
{DeCoursey}, C., {Egami}, E., {Pierel}, J. D.~R., {et~al.} 2025, \apj, 979,
  250, \dodoi{10.3847/1538-4357/ad8fab}

\bibitem[{{Dickinson} {et~al.}(2003){Dickinson}, {Giavalisco}, \& {GOODS
  Team}}]{dickinson2003}
{Dickinson}, M., {Giavalisco}, M., \& {GOODS Team}. 2003, in The Mass of
  Galaxies at Low and High Redshift, ed. R.~{Bender} \& A.~{Renzini}, 324,
  \dodoi{10.1007/10899892_78}

\bibitem[{{Dickinson} {et~al.}(2004){Dickinson}, {Stern}, {Giavalisco},
  {Ferguson}, {Tsvetanov}, {Chornock}, {Cristiani}, {Dawson}, {Dey},
  {Filippenko}, {Moustakas}, {Nonino}, {Papovich}, {Ravindranath}, {Riess},
  {Rosati}, {Spinrad}, \& {Vanzella}}]{dickinson2004}
{Dickinson}, M., {Stern}, D., {Giavalisco}, M., {et~al.} 2004, \apjl, 600, L99,
  \dodoi{10.1086/381119}

\bibitem[{{Egami} {et~al.}(2023){Egami}, {Sun}, {Alberts}, {Baum}, {Boyett},
  {Bunker}, {Cameron}, {Carniani}, {Charlot}, {Chen}, {Chevallard}, {Curti},
  {D'Eugenio}, {Danhaive}, {DeCoursey}, {Dudzeviciute}, {Eisenstein},
  {Hainline}, {Helton}, {Ji}, {Johnson}, {Kumari}, {Looser}, {Lyu}, {Ma},
  {Maiolino}, {Maseda}, {Nelson}, {Rawle}, {Rieke}, {Robertson}, {Sandles},
  {Shivaei}, {Smit}, {Suess}, {Tacchella}, {Uebler}, {Whitler}, {Williams},
  {Willmer}, {Willott}, {Witstok}, \& {de Graaff}}]{egami2023}
{Egami}, E., {Sun}, F., {Alberts}, S., {et~al.} 2023, {Complete NIRCam Grism
  Redshift Survey (CONGRESS)}, JWST Proposal. Cycle 2, ID. \#3577

\bibitem[{{Fei} {et~al.}(2025){Fei}, {Fujimoto}, {Naidu}, {Chisholm}, {Atek},
  {Brammer}, {Asada}, {Bromm}, {Furtak}, {Greene}, {Hsiao}, {Jeon}, {Kokorev},
  {Matthee}, {Natarajan}, {Richard}, {Saldana-Lopez}, {Schaerer}, {Volonteri},
  \& {Zitrin}}]{fei2025}
{Fei}, Q., {Fujimoto}, S., {Naidu}, R.~P., {et~al.} 2025, arXiv e-prints,
  arXiv:2509.20452, \dodoi{10.48550/arXiv.2509.20452}

\bibitem[{{Fei} {et~al.}(2026){Fei}, {Fujimoto}, {Brammer}, {Li}, {Ho},
  {Bromm}, {{\'A}lvarez-M{\'a}rquez}, {Asada}, {Barro}, {Colina}, {Dayal},
  {Finkelstein}, {Fynbo}, {Ginolfi}, {Inayoshi}, {Kokorev}, {Leung}, {Matthee},
  {Meyer}, {Naidu}, {Onoue}, {P{\'e}rez-Gonz{\'a}lez}, {Steinhardt},
  {Valentino}, {Walter}, {Xiao}, \& {Zhang}}]{fei2026}
{Fei}, Q., {Fujimoto}, S., {Brammer}, G., {et~al.} 2026, arXiv e-prints,
  arXiv:2602.12325.
\newblock \doarXiv{2602.12325}

\bibitem[{{French} {et~al.}(2020){French}, {Wevers}, {Law-Smith}, {Graur}, \&
  {Zabludoff}}]{french2020}
{French}, K.~D., {Wevers}, T., {Law-Smith}, J., {Graur}, O., \& {Zabludoff},
  A.~I. 2020, \ssr, 216, 32, \dodoi{10.1007/s11214-020-00657-y}

\bibitem[{{Fujimoto} {et~al.}(2022){Fujimoto}, {Brammer}, {Watson}, {Magdis},
  {Kokorev}, {Greve}, {Toft}, {Walter}, {Valiante}, {Ginolfi}, {Schneider},
  {Valentino}, {Colina}, {Vestergaard}, {Marques-Chaves}, {Fynbo}, {Krips},
  {Steinhardt}, {Cortzen}, {Rizzo}, \& {Oesch}}]{fujimoto2022}
{Fujimoto}, S., {Brammer}, G.~B., {Watson}, D., {et~al.} 2022, \nat, 604, 261,
  \dodoi{10.1038/s41586-022-04454-1}

\bibitem[{{Fujimoto} {et~al.}(2024){Fujimoto}, {Wang}, {Weaver}, {Kokorev},
  {Atek}, {Bezanson}, {Labbe}, {Brammer}, {Greene}, {Chemerynska}, {Dayal}, {de
  Graaff}, {Furtak}, {Oesch}, {Setton}, {Price}, {Miller}, {Williams},
  {Whitaker}, {Zitrin}, {Cutler}, {Leja}, {Pan}, {Coe}, {van Dokkum},
  {Feldmann}, {Fudamoto}, {Goulding}, {Khullar}, {Marchesini}, {Maseda},
  {Nanayakkara}, {Nelson}, {Smit}, {Stefanon}, \&
  {Weibel}}]{fujimoto2024uncover}
{Fujimoto}, S., {Wang}, B., {Weaver}, J.~R., {et~al.} 2024, \apj, 977, 250,
  \dodoi{10.3847/1538-4357/ad9027}

\bibitem[{{Fujimoto} {et~al.}(2025){Fujimoto}, {Ouchi}, {Kohno}, {Valentino},
  {Gim{\'e}nez-Arteaga}, {Brammer}, {Furtak}, {Kohandel}, {Oguri},
  {Pallottini}, {Richard}, {Zitrin}, {Bauer}, {Boylan-Kolchin},
  {Dessauges-Zavadsky}, {Egami}, {Finkelstein}, {Ma}, {Smail}, {Watson},
  {Hutchison}, {Rigby}, {Welch}, {Ao}, {Bradley}, {Caminha}, {Caputi},
  {Espada}, {Endsley}, {Fudamoto}, {Gonz{\'a}lez-L{\'o}pez}, {Hatsukade},
  {Koekemoer}, {Kokorev}, {Laporte}, {Lee}, {Magdis}, {Ono}, {Rizzo},
  {Shibuya}, {Shimasaku}, {Sun}, {Toft}, {Umehata}, {Wang}, \&
  {Yajima}}]{fujimoto2024}
{Fujimoto}, S., {Ouchi}, M., {Kohno}, K., {et~al.} 2025, Nature Astronomy, 9,
  1553, \dodoi{10.1038/s41550-025-02592-w}

\bibitem[{{Furtak} {et~al.}(2025){Furtak}, {Secunda}, {Greene}, {Zitrin},
  {Labb{\'e}}, {et~al.}}]{furtak2025}
{Furtak}, L.~J., {Secunda}, A.~R., {Greene}, J.~E., {et~al.} 2025, \aap, 698,
  A227, \dodoi{10.1051/0004-6361/202554110}

\bibitem[{{Gal-Yam}(2019)}]{galyam2019}
{Gal-Yam}, A. 2019, \araa, 57, 305, \dodoi{10.1146/annurev-astro-081817-051819}

\bibitem[{{Gezari}(2021)}]{gezari2021}
{Gezari}, S. 2021, \araa, 59, 21, \dodoi{10.1146/annurev-astro-111720-030029}

\bibitem[{{Greene} {et~al.}(2024){Greene}, {Labbe}, {Goulding}, {Furtak},
  {Chemerynska}, {Kokorev}, {Dayal}, {Volonteri}, {Williams}, {Wang}, {Setton},
  {Burgasser}, {Bezanson}, {Atek}, {Brammer}, {Cutler}, {Feldmann}, {Fujimoto},
  {Glazebrook}, {de Graaff}, {Khullar}, {Leja}, {Marchesini}, {Maseda},
  {Matthee}, {Miller}, {Naidu}, {Nanayakkara}, {Oesch}, {Pan}, {Papovich},
  {Price}, {van Dokkum}, {Weaver}, {Whitaker}, \& {Zitrin}}]{greene2024}
{Greene}, J.~E., {Labbe}, I., {Goulding}, A.~D., {et~al.} 2024, \apj, 964, 39,
  \dodoi{10.3847/1538-4357/ad1e5f}

\bibitem[{{Grogin} {et~al.}(2011){Grogin}, {Kocevski}, {Faber}, {Ferguson},
  {Koekemoer}, {Riess}, {Acquaviva}, {Alexander}, {Almaini}, {Ashby}, {Barden},
  {Bell}, {Bournaud}, {Brown}, {Caputi}, {Casertano}, {Cassata}, {Castellano},
  {Challis}, {Chary}, {Cheung}, {Cirasuolo}, {Conselice}, {Roshan Cooray},
  {Croton}, {Daddi}, {Dahlen}, {Dav{\'e}}, {de Mello}, {Dekel}, {Dickinson},
  {Dolch}, {Donley}, {Dunlop}, {Dutton}, {Elbaz}, {Fazio}, {Filippenko},
  {Finkelstein}, {Fontana}, {Gardner}, {Garnavich}, {Gawiser}, {Giavalisco},
  {Grazian}, {Guo}, {Hathi}, {H{\"a}ussler}, {Hopkins}, {Huang}, {Huang},
  {Jha}, {Kartaltepe}, {Kirshner}, {Koo}, {Lai}, {Lee}, {Li}, {Lotz}, {Lucas},
  {Madau}, {McCarthy}, {McGrath}, {McIntosh}, {McLure}, {Mobasher},
  {Moustakas}, {Mozena}, {Nandra}, {Newman}, {Niemi}, {Noeske}, {Papovich},
  {Pentericci}, {Pope}, {Primack}, {Rajan}, {Ravindranath}, {Reddy}, {Renzini},
  {Rix}, {Robaina}, {Rodney}, {Rosario}, {Rosati}, {Salimbeni}, {Scarlata},
  {Siana}, {Simard}, {Smidt}, {Somerville}, {Spinrad}, {Straughn}, {Strolger},
  {Telford}, {Teplitz}, {Trump}, {van der Wel}, {Villforth}, {Wechsler},
  {Weiner}, {Wiklind}, {Wild}, {Wilson}, {Wuyts}, {Yan}, \& {Yun}}]{grogin2011}
{Grogin}, N.~A., {Kocevski}, D.~D., {Faber}, S.~M., {et~al.} 2011, \apjs, 197,
  35, \dodoi{10.1088/0067-0049/197/2/35}

\bibitem[{{Guillochon} {et~al.}(2017){Guillochon}, {Parrent}, {Kelley}, \&
  {Margutti}}]{guilochon2017}
{Guillochon}, J., {Parrent}, J., {Kelley}, L.~Z., \& {Margutti}, R. 2017, \apj,
  835, 64, \dodoi{10.3847/1538-4357/835/1/64}

\bibitem[{{Harikane} {et~al.}(2023){Harikane}, {Zhang}, {Nakajima}, {Ouchi},
  {Isobe}, {Ono}, {Hatano}, {Xu}, \& {Umeda}}]{harikane2023agn}
{Harikane}, Y., {Zhang}, Y., {Nakajima}, K., {et~al.} 2023, \apj, 959, 39,
  \dodoi{10.3847/1538-4357/ad029e}

\bibitem[{{Hinkle} {et~al.}(2025){Hinkle}, {Shappee}, {Auchettl}, {Kochanek},
  {Neustadt}, {Polin}, {Strader}, {Holoien}, {Huber}, {Tucker}, {Ashall}, {de
  Jaeger}, {Desai}, {Do}, {Hoogendam}, \& {Payne}}]{hinkle2025}
{Hinkle}, J.~T., {Shappee}, B.~J., {Auchettl}, K., {et~al.} 2025, Science
  Advances, 11, eadt0074, \dodoi{10.1126/sciadv.adt0074}

\bibitem[{{H{\"o}nig}(2014)}]{honig2014}
{H{\"o}nig}, S.~F. 2014, \apjl, 784, L4, \dodoi{10.1088/2041-8205/784/1/L4}

\bibitem[{{Kajisawa} {et~al.}(2011){Kajisawa}, {Ichikawa}, {Tanaka}, {Yamada},
  {Akiyama}, {Suzuki}, {Tokoku}, {Katsuno Uchimoto}, {Konishi}, {Yoshikawa},
  {Nishimura}, {Omata}, {Ouchi}, {Iwata}, {Hamana}, \&
  {Onodera}}]{kajisawa2011}
{Kajisawa}, M., {Ichikawa}, T., {Tanaka}, I., {et~al.} 2011, \pasj, 63, 379,
  \dodoi{10.1093/pasj/63.sp2.S379}

\bibitem[{{Karmen} {et~al.}(2026){Karmen}, {Gezari}, {Norman}, \&
  {Guolo}}]{karmen2026}
{Karmen}, M., {Gezari}, S., {Norman}, C., \& {Guolo}, M. 2026, \apj, 1006, 20,
  \dodoi{10.3847/1538-4357/ae7a49}

\bibitem[{{Karmen} {et~al.}(2025){Karmen}, {Gezari}, {Lambrides}, {Akins},
  {Norman}, {Casey}, {Pierel}, {Coulter}, {Rest}, {Fox}, {Ajay}, {Allen},
  {Drakos}, {Fujimoto}, {Gomez}, {Gozaliasl}, {Ilbert}, {Kartaltepe},
  {Koekemoer}, {Lane}, {McCracken}, {Paquereau}, {Rhodes}, {Robertson},
  {Shuntov}, {Siebert}, {Toft}, {Wevers}, \& {Zenati}}]{karmen2025}
{Karmen}, M., {Gezari}, S., {Lambrides}, E., {et~al.} 2025, \apj, 990, 149,
  \dodoi{10.3847/1538-4357/adf216}

\bibitem[{{Lambrides} {et~al.}(2026){Lambrides}, {Hutchison}, {Larson},
  {Arrabal Haro}, {Papovich}, {et~al.}}]{lambrides2026}
{Lambrides}, E., {Hutchison}, T.~A., {Larson}, R.~L., {et~al.} 2026, arXiv
  e-prints, arXiv:2604.25991.
\newblock \doarXiv{2604.25991}

\bibitem[{{Leloudas} {et~al.}(2016){Leloudas}, {Fraser}, {Stone}, {van Velzen},
  {Jonker}, {Arcavi}, {Fremling}, {Maund}, {Smartt}, {Kr{\`\i}hler},
  {Miller-Jones}, {Vreeswijk}, {Gal-Yam}, {Mazzali}, {De Cia}, {Howell},
  {Inserra}, {Patat}, {de Ugarte Postigo}, {Yaron}, {Ashall}, {Bar},
  {Campbell}, {Chen}, {Childress}, {Elias-Rosa}, {Harmanen}, {Hosseinzadeh},
  {Johansson}, {Kangas}, {Kankare}, {Kim}, {Kuncarayakti}, {Lyman}, {Magee},
  {Maguire}, {Malesani}, {Mattila}, {McCully}, {Nicholl}, {Prentice},
  {Romero-Ca{\~n}izales}, {Schulze}, {Smith}, {Sollerman}, {Sullivan},
  {Tucker}, {Valenti}, {Wheeler}, \& {Young}}]{leloudas2016}
{Leloudas}, G., {Fraser}, M., {Stone}, N.~C., {et~al.} 2016, Nature Astronomy,
  1, 0002, \dodoi{10.1038/s41550-016-0002}

\bibitem[{{Leung} {et~al.}(2026{\natexlab{a}}){Leung}, {Eilers}, {Panagiotou},
  {Wolf}, {De}, {Weisenbach}, {Yue}, {Fan}, {Ishikawa}, {Kara}, {Krumpe},
  {Merloni}, {Simcoe}, {Wang}, \& {Yang}}]{leung2026}
{Leung}, G. C.~K., {Eilers}, A.-C., {Panagiotou}, C., {et~al.}
  2026{\natexlab{a}}, Nature Astronomy, \dodoi{10.1038/s41550-026-02897-4}

\bibitem[{{Leung} {et~al.}(2026{\natexlab{b}}){Leung}, {Eilers}, {Endsley},
  {Finkelstein}, {Bagley}, {Barro}, {Koekemoer}, {P{\'e}rez-Gonz{\'a}lez},
  {Pirzkal}, {Backhaus}, {Bulichi}, {Champagne}, {Chworowsky}, {Cleri},
  {Dickinson}, {Fan}, {Fujimoto}, {Grogin}, {Kirkpatrick}, {Kocevski},
  {Kokorev}, {Larson}, {Lucas}, {Pacucci}, {Papovich}, {Taylor}, {Villanueva},
  \& {Yang}}]{leung2026meow}
{Leung}, G.~C.~K., {Eilers}, A.-C., {Endsley}, R., {et~al.} 2026{\natexlab{b}},
  arXiv e-prints, \dodoi{10.48550/arXiv.2607.02666}

\bibitem[{{Li} {et~al.}(2024){Li}, {Ho}, {Ricci}, \& {Trakhtenbrot}}]{rli2024}
{Li}, R., {Ho}, L.~C., {Ricci}, C., \& {Trakhtenbrot}, B. 2024, \apj, 975, 50,
  \dodoi{10.3847/1538-4357/ad77a5}

\bibitem[{{Li} {et~al.}(2022){Li}, {Ho}, {Ricci}, {Trakhtenbrot}, {Arcavi},
  {Kara}, \& {Hiramatsu}}]{li2022}
{Li}, R., {Ho}, L.~C., {Ricci}, C., {et~al.} 2022, \apj, 933, 70,
  \dodoi{10.3847/1538-4357/ac714a}

\bibitem[{{Li} {et~al.}(2025){Li}, {Jiang}, {Wang}, {Shen}, {Qiao}, {Dai},
  {Luo}, {Li}, {Jin}, \& {Zhu}}]{wli2025}
{Li}, W., {Jiang}, N., {Wang}, T., {et~al.} 2025, arXiv e-prints,
  arXiv:2512.02147, \dodoi{10.48550/arXiv.2512.02147}

\bibitem[{{Lodato} \& {Rossi}(2011)}]{lodato2011}
{Lodato}, G., \& {Rossi}, E.~M. 2011, \mnras, 410, 359,
  \dodoi{10.1111/j.1365-2966.2010.17448.x}

\bibitem[{{MacLeod} {et~al.}(2010){MacLeod}, {Ivezi{\'c}}, {Kochanek},
  {Koz{\l}owski}, {Kelly}, {Bullock}, {Kimball}, {Sesar}, {Westman}, {Brooks},
  {Gibson}, {Becker}, \& {de Vries}}]{macleod2010}
{MacLeod}, C.~L., {Ivezi{\'c}}, {\v{Z}}., {Kochanek}, C.~S., {et~al.} 2010,
  \apj, 721, 1014, \dodoi{10.1088/0004-637X/721/2/1014}

\bibitem[{{MacLeod} {et~al.}(2019){MacLeod}, {Green}, {Anderson}, {Bruce},
  {Eracleous}, {Graham}, {Homan}, {Lawrence}, {LeBleu}, {Ross}, {Ruan},
  {Runnoe}, {Stern}, {Burgett}, {Chambers}, {Kaiser}, {Magnier}, \&
  {Metcalfe}}]{macleod2019}
{MacLeod}, C.~L., {Green}, P.~J., {Anderson}, S.~F., {et~al.} 2019, \apj, 874,
  8, \dodoi{10.3847/1538-4357/ab05e2}

\bibitem[{{Magnelli} {et~al.}(2011){Magnelli}, {Elbaz}, {Chary}, {Dickinson},
  {Le Borgne}, {Frayer}, \& {Willmer}}]{magnelli2011}
{Magnelli}, B., {Elbaz}, D., {Chary}, R.~R., {et~al.} 2011, \aap, 528, A35,
  \dodoi{10.1051/0004-6361/200913941}

\bibitem[{{Maiolino} {et~al.}(2024){Maiolino}, {Scholtz}, {Curtis-Lake},
  {Carniani}, {Baker}, {de Graaff}, {Tacchella}, {{\"U}bler}, {D'Eugenio},
  {Witstok}, {Curti}, {Arribas}, {Bunker}, {Charlot}, {Chevallard},
  {Eisenstein}, {Egami}, {Ji}, {Jones}, {Lyu}, {Rawle}, {Robertson},
  {Rujopakarn}, {Perna}, {Sun}, {Venturi}, {Williams}, \&
  {Willott}}]{maiolino2024jades}
{Maiolino}, R., {Scholtz}, J., {Curtis-Lake}, E., {et~al.} 2024, \aap, 691,
  A145, \dodoi{10.1051/0004-6361/202347640}

\bibitem[{{Matthee} {et~al.}(2024){Matthee}, {Naidu}, {Brammer}, {Chisholm},
  {Eilers}, {Goulding}, {Greene}, {Kashino}, {Labbe}, {Lilly}, {Mackenzie},
  {Oesch}, {Weibel}, {Wuyts}, {Xiao}, {Bordoloi}, {Bouwens}, {van Dokkum},
  {Illingworth}, {Kramarenko}, {Maseda}, {Mason}, {Meyer}, {Nelson}, {Reddy},
  {Shivaei}, {Simcoe}, \& {Yue}}]{matthee2024}
{Matthee}, J., {Naidu}, R.~P., {Brammer}, G., {et~al.} 2024, \apj, 963, 129,
  \dodoi{10.3847/1538-4357/ad2345}

\bibitem[{{Moriya} {et~al.}(2025){Moriya}, {Coulter}, {DeCoursey}, {Pierel},
  {Hainline}, {Siebert}, {Rest}, {Egami}, {Gomez}, {Quimby}, {Fox}, {Engesser},
  {Sun}, {Chen}, {Zenati}, {Gezari}, {Joshi}, {Shahbandeh}, {Strolger}, {Wang},
  {Alberts}, {Bhatawdekar}, {Bunker}, {Rinaldi}, {Robertson}, \&
  {Tacchella}}]{moriya2025}
{Moriya}, T.~J., {Coulter}, D.~A., {DeCoursey}, C., {et~al.} 2025, \pasj, 77,
  851, \dodoi{10.1093/pasj/psaf052}

\bibitem[{{Mowla} {et~al.}(2024){Mowla}, {Iyer}, {Asada}, {Desprez}, {Tan},
  {Martis}, {Sarrouh}, {Strait}, {Abraham}, {Brada{\v{c}}}, {Brammer},
  {Muzzin}, {Pacifici}, {Ravindranath}, {Sawicki}, {Willott},
  {Estrada-Carpenter}, {Jahan}, {Noirot}, {Matharu}, {Rihtar{\v{s}}i{\v{c}}},
  \& {Zabl}}]{mowla2024}
{Mowla}, L., {Iyer}, K., {Asada}, Y., {et~al.} 2024, \nat, 636, 332,
  \dodoi{10.1038/s41586-024-08293-0}

\bibitem[{{Naidu} {et~al.}(2025){Naidu}, {Chisholm}, {Matthee}, {Bordoloi},
  {Brammer}, {De}, {Eilers}, {Fudamoto}, {Furtak}, {Greene}, {Kara},
  {Kramarenko}, {Mackenzie}, {Marchesini}, {Oesch}, {Shen}, {Vogelsberger}, \&
  {Yue}}]{naidu2025a}
{Naidu}, R., {Chisholm}, J., {Matthee}, J., {et~al.} 2025, {How I wonder what
  you are - do JWST's Little Red Dots twinkle? Testing broad-line and continuum
  variability on week, month, and six-month timescales}, JWST Proposal. Cycle
  4, ID. \#7404

\bibitem[{{Nicholl} {et~al.}(2016){Nicholl}, {Berger}, {Smartt}, {Margutti},
  {Kamble}, {Alexander}, {Chen}, {Inserra}, {Arcavi}, {Blanchard}, {Cartier},
  {Chambers}, {Childress}, {Chornock}, {Cowperthwaite}, {Drout}, {Flewelling},
  {Fraser}, {Gal-Yam}, {Galbany}, {Harmanen}, {Holoien}, {Hosseinzadeh},
  {Howell}, {Huber}, {Jerkstrand}, {Kankare}, {Kochanek}, {Lin}, {Lunnan},
  {Magnier}, {Maguire}, {McCully}, {McDonald}, {Metzger}, {Milisavljevic},
  {Mitra}, {Reynolds}, {Saario}, {Shappee}, {Smith}, {Valenti}, {Villar},
  {Waters}, \& {Young}}]{nicholl2016}
{Nicholl}, M., {Berger}, E., {Smartt}, S.~J., {et~al.} 2016, \apj, 826, 39,
  \dodoi{10.3847/0004-637X/826/1/39}

\bibitem[{{Oesch} {et~al.}(2023){Oesch}, {Brammer}, {Naidu}, {Bouwens},
  {Chisholm}, {Illingworth}, {Matthee}, {Nelson}, {Qin}, {Reddy}, {Shapley},
  {Shivaei}, {van Dokkum}, {Weibel}, {Whitaker}, {Wuyts}, {Covelo-Paz},
  {Endsley}, {Fudamoto}, {Giovinazzo}, {Herard-Demanche}, {Kerutt},
  {Kramarenko}, {Labbe}, {Leonova}, {Lin}, {Magee}, {Marchesini}, {Maseda},
  {Mason}, {Matharu}, {Meyer}, {Neufeld}, {Prieto Lyon}, {Schaerer}, {Sharma},
  {Shuntov}, {Smit}, {Stefanon}, {Wyithe}, \& {Xiao}}]{oesch2023}
{Oesch}, P.~A., {Brammer}, G., {Naidu}, R.~P., {et~al.} 2023, \mnras, 525,
  2864, \dodoi{10.1093/mnras/stad2411}

\bibitem[{{Phinney}(1989)}]{phinney1989}
{Phinney}, E.~S. 1989, in IAU Symposium, Vol. 136, The Center of the Galaxy,
  ed. M.~{Morris}, 543

\bibitem[{{Planck Collaboration} {et~al.}(2014){Planck Collaboration}, {Ade},
  {Aghanim}, {Armitage-Caplan}, {Arnaud}, {Ashdown}, {Atrio-Barandela},
  {Aumont}, {Baccigalupi}, {Banday}, \& et~al.}]{planck2014}
{Planck Collaboration}, {Ade}, P.~A.~R., {Aghanim}, N., {et~al.} 2014, \aap,
  571, A16, \dodoi{10.1051/0004-6361/201321591}

\bibitem[{{Rees}(1988)}]{rees1988}
{Rees}, M.~J. 1988, \nat, 333, 523, \dodoi{10.1038/333523a0}

\bibitem[{{Stefanon} {et~al.}(2021){Stefanon}, {Bouwens}, {Labb{\'e}},
  {Illingworth}, {Gonzalez}, \& {Oesch}}]{stefanon2021}
{Stefanon}, M., {Bouwens}, R.~J., {Labb{\'e}}, I., {et~al.} 2021, \apj, 922,
  29, \dodoi{10.3847/1538-4357/ac1bb6}

\bibitem[{{Stone} \& {Metzger}(2016)}]{stone2016b}
{Stone}, N.~C., \& {Metzger}, B.~D. 2016, \mnras, 455, 859,
  \dodoi{10.1093/mnras/stv2281}

\bibitem[{{Stone} \& {van Velzen}(2016)}]{stone2016a}
{Stone}, N.~C., \& {van Velzen}, S. 2016, \apjl, 825, L14,
  \dodoi{10.3847/2041-8205/825/1/L14}

\bibitem[{{van Velzen} {et~al.}(2016){van Velzen}, {Mendez}, {Krolik}, \&
  {Gorjian}}]{van-velzen2016}
{van Velzen}, S., {Mendez}, A.~J., {Krolik}, J.~H., \& {Gorjian}, V. 2016,
  \apj, 829, 19, \dodoi{10.3847/0004-637X/829/1/19}

\bibitem[{{van Velzen} {et~al.}(2021){van Velzen}, {Gezari}, {Hammerstein},
  {Roth}, {Frederick}, {Ward}, {Hung}, {Cenko}, {Stein}, {Perley}, {Taggart},
  {Foley}, {Sollerman}, {Blagorodnova}, {Andreoni}, {Bellm}, {Brinnel}, {De},
  {Dekany}, {Feeney}, {Fremling}, {Giomi}, {Golkhou}, {Graham}, {Ho},
  {Kasliwal}, {Kilpatrick}, {Kulkarni}, {Kupfer}, {Laher}, {Mahabal}, {Masci},
  {Miller}, {Nordin}, {Riddle}, {Rusholme}, {van Santen}, {Sharma}, {Shupe}, \&
  {Soumagnac}}]{van-velzen2021}
{van Velzen}, S., {Gezari}, S., {Hammerstein}, E., {et~al.} 2021, \apj, 908, 4,
  \dodoi{10.3847/1538-4357/abc258}

\bibitem[{{Vanzella} {et~al.}(2023){Vanzella}, {Claeyssens}, {Welch}, {Adamo},
  {Coe}, {Diego}, {Mahler}, {Khullar}, {Kokorev}, {Oguri}, {Ravindranath},
  {Furtak}, {Hsiao}, {Abdurro'uf}, {Mandelker}, {Brammer}, {Bradley},
  {Brada{\v{c}}}, {Conselice}, {Dayal}, {Nonino}, {Andrade-Santos},
  {Windhorst}, {Pirzkal}, {Sharon}, {de Mink}, {Fujimoto}, {Zitrin},
  {Eldridge}, \& {Norman}}]{vanzella2023}
{Vanzella}, E., {Claeyssens}, A., {Welch}, B., {et~al.} 2023, \apj, 945, 53,
  \dodoi{10.3847/1538-4357/acb59a}

\bibitem[{{Wang} {et~al.}(2021){Wang}, {Yang}, {Fan}, {Hennawi}, {Barth},
  {Banados}, {Bian}, {Boutsia}, {Connor}, {Davies}, {Decarli}, {Eilers},
  {Farina}, {Green}, {Jiang}, {Li}, {Mazzucchelli}, {Nanni}, {Schindler},
  {Venemans}, {Walter}, {Wu}, \& {Yue}}]{wang2021}
{Wang}, F., {Yang}, J., {Fan}, X., {et~al.} 2021, \apjl, 907, L1,
  \dodoi{10.3847/2041-8213/abd8c6}

\bibitem[{{Yoshii} {et~al.}(2014){Yoshii}, {Kobayashi}, {Minezaki}, {Koshida},
  \& {Peterson}}]{yoshii2014}
{Yoshii}, Y., {Kobayashi}, Y., {Minezaki}, T., {Koshida}, S., \& {Peterson},
  B.~A. 2014, \apjl, 784, L11, \dodoi{10.1088/2041-8205/784/1/L11}

\bibitem[{{Zhu} {et~al.}(2026){Zhu}, {Xu}, {Jiang}, {Jiang}, {Wang}, {Yao},
  {Chornock}, {Hammerstein}, {et~al.}}]{zhu2026}
{Zhu}, J., {Xu}, Z., {Jiang}, N., {et~al.} 2026, \apjl, 1004, L10,
  \dodoi{10.3847/2041-8213/ae710d}

\end{thebibliography}

\end{document}